\documentclass[aps,prc,twocolumn,showpacs,preprintnumbers,amsmath,superscriptaddress,amssymb,floats,nofootinbib]{revtex4} 

\usepackage{graphicx}
\usepackage{dcolumn}
\begin{document} 
 
\def\Journal#1#2#3#4{{#1} {\bf #2}, #3 (#4)} 
\def\NCA{\rm Nuovo Cimento}
\def\NCL{\rm Lett. Nuovo Cimento}
\def\NIM{\rm Nucl. Instrum. Methods}
\def\NIMA{{\rm Nucl. Instrum. Methods} A}
\def\NPA{{\rm Nucl. Phys.} A}
\def\NPB{{\rm Nucl. Phys.} B}
\def\PLB{{\rm Phys. Lett.}  B}
\def\PRL{\rm Phys. Rev. Lett.}
\def\PRD{{\rm Phys. Rev.} D}
\def\PRC{{\rm Phys. Rev.} C}
\def\EPA{{\rm Eur. Phys. J.} A}
\def\EPC{{\rm Eur. Phys. J.} C}
\def\ZPC{{\rm Z. Phys.} C}
\def\ZPA{{\rm Z. Phys.} A}
\def\JPG{{\rm J. Phys.} G}
\title{Cross Section Measurements of Charged Pion Photoproduction\\ 
 in Hydrogen and Deuterium from 1.1 to 5.5 GeV} 
\author{L.~Y.~Zhu} 
\affiliation{Massachusetts Institute of Technology, Cambridge, MA 
02139, USA} 
\author{J.~Arrington} 
\affiliation{Argonne National Laboratory, Argonne, IL 60439, USA} 
\author{T.~Averett} 
\affiliation{College of William and Mary, Williamsburg, VA 23185, USA} 
\affiliation{Thomas Jefferson National Accelerator Facility, Newport 
News, VA 23606, USA} 
\author{E.~Beise} 
\affiliation{University of Maryland, College Park, MD 20742, USA} 
\author{J.~Calarco} 
\affiliation{University of New Hampshire, Durham, NH 03824, USA} 
\author{T.~Chang} 
\affiliation{University of Illinois, Urbana, IL 61801, USA} 
\author{J.~P.~Chen} 
\affiliation{Thomas Jefferson National Accelerator Facility, Newport 
News, VA 23606, USA} 
\author{E.~Chudakov} 
\affiliation{Thomas Jefferson National Accelerator Facility, Newport 
News, VA 23606, USA} 
\author{M.~Coman} 
\affiliation{Florida International University, Miami, FL 33199, USA} 
\author{B.~Clasie} 
\affiliation{Massachusetts Institute of Technology, Cambridge, MA 
02139, USA} 
\author{C.~Crawford} 
\affiliation{Massachusetts Institute of Technology, Cambridge, MA 
02139, USA} 
\author{S.~Dieterich} 
\affiliation{Rutgers University, New Brunswick, NJ 08903, USA} 
\author{F.~Dohrmann} 
\affiliation{Argonne National Laboratory, Argonne, IL 60439, USA} 
\author{D.~Dutta} 
\affiliation{Massachusetts Institute of Technology, Cambridge, MA 
02139, USA} 
\author{K.~Fissum} 
\affiliation{Lund University, S-221 00 Lund, Sweden} 
\author{S.~Frullani} 
\affiliation{Istituto Nazionale di Fisica Nucleare, Sezione Sanit\`{a}, 00161 Roma, Italy} 
\author{H.~Gao} 
\affiliation{Massachusetts Institute of Technology, Cambridge, MA 
02139, USA} 
\affiliation{Duke University, Durham, NC 27708, USA} 
\author{R.~Gilman} 
\affiliation{Thomas Jefferson National Accelerator Facility, Newport 
News, VA 23606, USA} 
\affiliation{Rutgers University, New Brunswick, NJ 08903, USA} 
\author{C.~Glashausser} 
\affiliation{Rutgers University, New Brunswick, NJ 08903, USA} 
\author{J.~Gomez} 
\affiliation{Thomas Jefferson National Accelerator Facility, Newport 
News, VA 23606, USA} 
\author{K.~Hafidi} 
\affiliation{Argonne National Laboratory, Argonne, IL 60439, USA} 
\author{O.~Hansen} 
\affiliation{Thomas Jefferson National Accelerator Facility, Newport 
News, VA 23606, USA} 
\author{D.~W.~Higinbotham} 
\affiliation{Massachusetts Institute of Technology, Cambridge, MA 
02139, USA} 
\author{R.~J.~Holt} 
\affiliation{Argonne National Laboratory, Argonne, IL 60439, USA} 
\author{C.~W.~de Jager} 
\affiliation{Thomas Jefferson National Accelerator Facility, Newport 
News, VA 23606, USA} 
\author{X.~Jiang} 
\affiliation{Rutgers University, New Brunswick, NJ 08903, USA} 
\author{E.~Kinney} 
\affiliation{University of Colorado, Boulder, CO 80302, USA} 
\author{K.~Kramer} 
\affiliation{College of William and Mary, Williamsburg, VA 23185, USA} 
\author{G.~Kumbartzki} 
\affiliation{Rutgers University, New Brunswick, NJ 08903, USA} 
\author{J.~LeRose} 
\affiliation{Thomas Jefferson National Accelerator Facility, Newport 
News, VA 23606, USA} 
\author{N.~Liyanage} 
\affiliation{Thomas Jefferson National Accelerator Facility, Newport 
News, VA 23606, USA} 
\author{D.~Mack} 
\affiliation{Thomas Jefferson National Accelerator Facility, Newport 
News, VA 23606, USA} 
\author{P.~Markowitz} 
\affiliation{Florida International University, Miami, FL 33199, USA} 
\author{K.~McCormick} 
\affiliation{Rutgers University, New Brunswick, NJ 08903, USA} 
\author{D.~Meekins} 
\affiliation{Thomas Jefferson National Accelerator Facility, Newport 
News, VA 23606, USA} 
\author{Z.-E.~Meziani} 
\affiliation{Temple University, Philadelphia, PA 19122, USA} 
\author{R.~Michaels} 
\affiliation{Thomas Jefferson National Accelerator Facility, Newport 
News, VA 23606, USA} 
\author{J.~Mitchell} 
\affiliation{Thomas Jefferson National Accelerator Facility, Newport 
News, VA 23606, USA} 
\author{S.~Nanda} 
\affiliation{Thomas Jefferson National Accelerator Facility, Newport 
News, VA 23606, USA} 
\author{D.~Potterveld} 
\affiliation{Argonne National Laboratory, Argonne, IL 60439, USA} 
\author{R.~Ransome} 
\affiliation{Rutgers University, New Brunswick, NJ 08903, USA} 
\author{P.~E.~Reimer} 
\affiliation{Argonne National Laboratory, Argonne, IL 60439, USA} 
\author{B.~Reitz} 
\affiliation{Thomas Jefferson National Accelerator Facility, Newport 
News, VA 23606, USA} 
\author{A.~Saha} 
\affiliation{Thomas Jefferson National Accelerator Facility, Newport 
News, VA 23606, USA} 
\author{E.~C.~Schulte} 
\affiliation{Argonne National Laboratory, Argonne, IL 60439, USA} 
\affiliation{University of Illinois, Urbana, IL 61801, USA} 
\author{J.~Seely} 
\affiliation{Massachusetts Institute of Technology, Cambridge, MA 
02139, USA} 
\author{S.~\v{S}irca} 
\affiliation{Massachusetts Institute of Technology, Cambridge, MA 
02139, USA} 
\author{S.~Strauch} 
\affiliation{Rutgers University, New Brunswick, NJ 08903, USA} 
\author{V.~Sulkosky} 
\affiliation{College of William and Mary, Williamsburg, VA 23185, USA} 
\author{B.~Vlahovic} 
\affiliation{North Carolina Central University, Durham, NC 27707, USA} 
\author{L.~B.~Weinstein} 
\affiliation{Old Dominion University, Norfolk, VA 23529, USA} 
\author{K.~Wijesooriya} 
\affiliation{Argonne National Laboratory, Argonne, IL 60439, USA} 
\author{C.~Williamson} 
\affiliation{Massachusetts Institute of Technology, Cambridge, MA 
02139, USA} 
\author{B.~Wojtsekhowski} 
\affiliation{Thomas Jefferson National Accelerator Facility, Newport 
News, VA 23606, USA} 
\author{H.~Xiang} 
\affiliation{Massachusetts Institute of Technology, Cambridge, MA 
02139, USA} 
\author{F.~Xiong} 
\affiliation{Massachusetts Institute of Technology, Cambridge, MA 
02139, USA} 
\author{W.~Xu} 
\affiliation{Massachusetts Institute of Technology, Cambridge, MA 
02139, USA} 
\author{J.~Zeng} 
\affiliation{University of Georgia, Athens, GA 30601, USA} 
\author{X.~Zheng} 
\affiliation{Massachusetts Institute of Technology, Cambridge, MA 
02139, USA} 
 
\collaboration{Jefferson Lab Hall A Collaboration and Jefferson Lab E94-104 Collaboration} 
\noaffiliation 
\date{\today} 
 
\begin{abstract} 
The differential cross sections for the $\gamma n \rightarrow \pi^- p$ and the $\gamma p \rightarrow \pi^+ n$ processes were measured at Jefferson Lab. The photon energies ranged from 1.1 to 5.5 GeV, corresponding to center-of-mass energies from 1.7 to 3.4 GeV. The pion center-of-mass angles varied from  50$^\circ$ to 110$^\circ$. The $\pi^-$ and $\pi^+$ photoproduction data both exhibit a global scaling behavior at high energies and high transverse momenta, consistent with the constituent counting rule prediction and the existing $\pi^+$ data.  The data suggest possible substructure of the scaling behavior, which might be oscillations around the scaling value. The data show an enhancement in the scaled cross section at center-of-mass energy near 2.2 GeV. The differential cross section ratios $\frac{d\sigma/dt(\gamma n \rightarrow \pi^- p)}{d\sigma/dt(\gamma p \rightarrow \pi^+ n)}$ at high energies and high transverse momenta can be described by calculations based on one-hard-gluon-exchange diagrams.  
 
\end{abstract} 
 

\maketitle 
\section{Introduction} 
Quantum ChromoDynamics (QCD), a fundamental theory for describing the strong interaction, is not amenable to analytical solutions in the nonperturbative region. Some dynamical models must be developed. Meson-exchange models in terms of the nucleon-meson degrees of freedom describe nuclear physics data well at low energy, and perturbative QCD (pQCD) in terms of the quark-gluon degrees of freedom succeeds in explaining many measurements at high energy. But little is known about the transition between these two regions. Testing the constituent counting rule for the exclusive reactions is one way to study the transition of the degrees of freedom. 
 
The constituent counting rule establishes a direct connection between the quark-gluon degrees of freedom and the energy dependence of the differential cross section for exclusive processes at fixed center-of-mass angles. This rule was first derived from simple dimensional counting~\cite{brodsky73,brodsky75,matveev73} and was later confirmed in a short-distance pQCD approach~\cite{lepage80}. It is consistent with many exclusive measurements~\cite{land73,white94,schulte01,hallb04,hallb04b,anderson76}. However, there are still many puzzles. This rule begins to agree with experimental data at photon energies as low as 1 GeV~\cite{schulte01}, whereas pQCD is not expected to be valid at such low energies. The hadron helicity 
conservation rule~\cite{hhc_brodsky}, another outcome of the same short-distance pQCD framework, does not agree with data in the same energy and momentum transfer region~\cite{krishni01,krishni02}. There are a few anomalies beyond the constituent counting rule in the extensively studied $pp$ scattering process~\cite{crabb,hendry,ct5}. Recently, the parton orbital angular momentum 
has been found to play a non-negligible role in the exclusive reactions, which may explain hadron helicity nonconservation and other polarization measurements~\cite{gousset,belisky}. 
 
Single pion photoproduction, $\gamma N \rightarrow \pi N $, is a relatively simple process for studying the strong interaction. It has larger cross sections at high energy than other exclusive channels due to its slower decrease with energy, i.e. $d\sigma /dt \sim s^{-7}$. One can also form the differential cross section ratio $\frac{d\sigma/dt(\gamma n \rightarrow \pi^- p)}{d\sigma/dt(\gamma p \rightarrow \pi^+ n)}$. The ratio is amenable to theoretical predictions  since many factors may cancel out in leading order.
 
This paper focuses on extracting the differential cross sections for the single charged pion photoproduction processes, $\gamma p \rightarrow \pi^+ n$ and $\gamma n \rightarrow \pi^- p$. This is one major goal of experiment E94-104~\cite{E94-104} carried out at Jefferson Lab (JLab). The photon beam energy ranged from 1.1 to 5.5 GeV. The pion center-of-mass angles varied from $50^\circ$ to $110^\circ$. The results at 90$^\circ$ have already been published~\cite{prl_90}, which are also updated in this paper. The experiment E94-104 also measured $\pi^-$ photoproduction with a helium target, providing the first nuclear transparency data for this process~\cite{prc_he}. 
 
This paper is organized as follows. Section II introduces the theoretical and experimental background for JLab experiment E94-104. Section III describes this experiment at JLab Hall A. Section IV presents the data analysis procedure to extract the differential cross sections. Section V discusses the results. Section VI and VII are outlook and acknowledgments. 
\section{Theoretical and Experimental Background} 
The constituent counting rule is also called the dimensional scaling rule. It states that 
\begin{equation} 
(d\sigma/dt)_{AB \rightarrow CD} \sim s^{-(n-2)} f(\theta_{c.m.})  
\label{eq:ccr} 
\end{equation} 
for an exclusive two-body reaction $AB \rightarrow CD$ when $ s \rightarrow \infty$. Here $s$ and $t$ are the Mandelstam variables and $n$ is the total number of elementary fields (quarks, leptons or photons) which carry finite fractions of particle momentum. It states that at fixed center-of-mass angle $\theta_{c.m.}$ and large $s$, $(d\sigma/dt)_{pp\rightarrow pp}\sim s^{-10}$, $(d\sigma/dt)_{\pi p\rightarrow \pi p}\sim s^{-8}$, $(d\sigma/dt)_{\gamma d \rightarrow pn}\sim s^{-11}$ and $(d\sigma/dt)_{\gamma N \rightarrow \pi N}\sim s^{-7}$. The constituent counting rule implies something of fundamental importance: the quark has not only a mathematical existence, giving current algebra, Bjorken scaling and the hadron spectrum, but a dynamical existence as well~\cite{brodsky75}. 
 
The constituent counting rule was originally derived from simple dimensional counting by Brodsky and Farrar~\cite{brodsky73}, and simultaneously by Matveev {\it el al.}~\cite{matveev73} in 1973.  
Brodsky and Farrar also examined the required conditions for the simple dimensional derivation: \\ 
(a) the effective replacement of the composite hadron by constituents carrying finite fractions of the hadron momentum, \\ 
(b) the absence of any mass scale in the amplitude or binding corrections.\\ 
They showed that both condition (a) and (b) are natural features of renormalizable field theories, with certain dynamical assumptions concerning the nature of the wave function, the absence of infrared effects, and the accumulation of logarithms~\cite{brodsky75}. 
 
Later in 1980, Lepage and Brodsky showed that the constituent counting rule can be reproduced within a short-distance pQCD approach~\cite{lepage80}, up to calculable powers of the strong coupling constant. The energy dependence of the strong coupling constant is small at high energy, and some recent $\tau$ decay data also suggest the freezing of the coupling constant at low energy~\cite{brodsky_alpha}.

Another outcome of this approach is hadron helicity 
conservation~\cite{hhc_brodsky}:  
\begin{equation} 
h_A+h_B = h_C + h_D, 
\end{equation}  
which leads to strong correlations between the final state helicities. 
The above results came from the calculation of an enormous number of connected tree diagrams for hard subprocesses without considering the parton orbital angular momentum, while the soft subprocesses, such as Landshoff diagrams~\cite{landshoff} were suppressed in leading order for example due to gluon radiation. 
 
The scaling behavior of the differential cross section predicted by the constituent counting rule can also be described by string theory~\cite{string} and other phenomenological models. For the deuteron photodisintegration process as an example, the Quark-Gluon String (QGS) model~\cite{QGS} and the Hard Rescattering Mechanism (HRM)~\cite{HRM} describe fairly well both the energy dependence and the asymmetric angular distribution of the data~\cite{schulte01,hallb04,schulte_th,schulte02}. The Reduced Nuclear Amplitudes (RNA)~\cite{RNA} and the Asymptotic Meson Exchange Calculation (AMEC)~\cite{AMEC} can describe the scaling behavior at $\theta_{c.m.}=90^\circ$~\cite{schulte01,hallb04,schulte_th}.   

On the experimental side, the constituent counting rule is consistent with data for many exclusive processes, such as $pp$ elastic scattering~\cite{land73}, hadron-hadron elastic scattering~\cite{white94} and deuteron photodisintegration~\cite{schulte01,hallb04,schulte_th}. The fitted power of $\frac{1}{s}$ from the $pp$ elastic scattering data with $s > 15 \ {\rm GeV}^2$ and $|t| > 2.5 \ {\rm GeV}^2$  is equal to $9.7 \pm 0.5$~\cite{land73}, consistent with 10 as predicted by the constituent counting rule. Eight meson-baryon and two baryon-baryon exclusive reactions at $\theta_{c.m.} =90^\circ$ were measured at the AGS (the Alternate Gradient Synchrotron at BNL) with beam momenta of 5.9 GeV/c and 9.9 GeV/c. The fitted powers of $\frac{1}{s}$ are also consistent with the constituent counting rule predictions, i.e. 8 for the meson-baryon reactions and 10 for the baryon-baryon reactions, except for one reaction: $\pi^- p \rightarrow \pi^+ \Delta^-$~\cite{white94}. Deuteron photodisintegration is another process that exhibits scaling behavior of the differential cross sections~\cite{schulte01,hallb04,schulte_th}. The onset of scaling in photon energy for deuteron photodisintegration depends greatly on the center-of-mass angles. The corresponding threshold of the proton transverse momentum, $P_T=\sqrt{\frac{1}{2}M_d E_\gamma {\rm sin}^2\theta_{c.m.}}$, is above about 1.1 GeV/c for the proton angle between 30$^\circ$ and 150$^\circ$~\cite{hallb04b}.

 
Despite the theoretical and experimental support for the constituent counting rule, there remain some puzzles and anomalies. First of all, it is surprising to see the onset of scaling at transverse momentum as low as about 1.1 GeV/c~\cite{schulte01,hallb04b}, such as in the photodisintegration data. The applicability of pQCD to exclusive processes remains controversial in the GeV region. The pQCD calculation fails to predict the magnitude of some fundamental quantities, such as the proton magnetic form factor $G_M^p$ with $Q^2$ up to 30 (GeV/c)$^2$~\cite{isgur}. Hadron helicity conservation, another consequence of pQCD (this statement is currently under debate~\cite{ralston02}), tends not to agree with polarization measurements, such as those from JLab for the photodisintegration process $d(\vec{\gamma},\vec{p})n$ up to 2.4 GeV~\cite{krishni01} and neutral pion photoproduction $p(\vec{\gamma},\vec{p})\pi^0$ up to 4.1 GeV~\cite{krishni02}. 
Although contributions from nonzero parton orbital angular momentum 
are power suppressed as shown by Lepage and Brodsky~\cite{lepage80}, they 
could break the hadron helicity 
conservation rule~\cite{gousset}. 
Orbital angular momentum could also lead to asymptotic scaling of the 
proton form factor ratio: $F_2(Q^2)/F_1(Q^2)\sim ({\rm log}^2Q^2/ \Lambda^2)/Q^2$ with 0.2 GeV$\le \Lambda \le $0.4 GeV based on an explicit pQCD calculation~\cite{belisky} or $F_2(Q^2)/F_1(Q^2)\sim 1/\sqrt{Q^2}$~\cite{ralston02,miller} that agrees with the JLab proton form factor data~\cite{gep}.
A recent nonperturbative analysis~\cite{brodsky_ff} of the hadronic 
form factors based on light-front wave functions, and a model with an intrinsic (quark-like) structure and a meson cloud~\cite{iachello} also describes the 
JLab proton form factor data~\cite{gep} well.
 
Furthermore, several striking anomalies have been observed in $pp$ scattering. 
One is the very large spin-spin correlation. The ratio of $(d \sigma/dt)_{\uparrow \uparrow}/(d \sigma/dt)_{\uparrow \downarrow}$ with spin normal to the scattering plane can reach 4 in $pp$ elastic scattering at $\theta_{c.m.}=90^\circ$~\cite{crabb}. Next is the oscillation of the differential cross section $d\sigma /dt$ around the scaling value in $pp$ elastic scattering~\cite{hendry}. The third is the anomalous energy dependence of nuclear transparency of the $A(p,2p)$ process~\cite{ct5}.  
 
There exist different theoretical attempts to describe the anomalies.  
One example is the interference between two types of subprocesses, the short-distance hard subprocesses and the long-distance soft (Landshoff) subprocesses~\cite{ralston}. Another example is the interference between the pQCD background and two $J=L=S=1, B=2$ resonance structures associated with the strangeness and charm production thresholds~\cite{brodsky_de}.  

Recently, more mechanisms were under discussion that can lead to the deviation from the scaling. For example, the generalized constituent counting rule~\cite{ji03,brodsky03,dutta04} including the parton orbital angular momentum and the restricted locality of quark-hadron duality~\cite{zhao03}.
 
Including the nonzero parton orbital angular momentum may change the expression of the constituent counting rule. Based on a hadronic light-cone wave function involving parton orbital angular momentum, which has been used to describe the JLab proton form factor data, a generalized constituent counting rule~\cite{ji03} can be derived for hard exclusive processes by counting the soft mass dimensions of scattering amplitudes. This generalized constituent counting rule with parton orbital angular momentum dependence can be also derived~\cite{brodsky03} in a non-perturbative method, which does not rely on pQCD.
According to the generalized constituent counting rule, the fixed-angle scattering cross section behaves like
\begin{equation}
\Delta \sigma \sim s^{-1-\sum_H{(n_H+|l_{zH}|-1)}},
\end{equation} 
for the exclusive reaction $A+B\rightarrow C+D+...$, where $n_H$ is the number of elementary fields in involved hadron $H$ and $\Delta \sigma$ contains only angular variables. For parton orbital angular momentum $l_{zH}=0$, this is just the traditional constituent counting rule. As a result, the helicity-flip amplitudes for the $p p \rightarrow pp $ process were predicted to scale as $s^{-9/2}$ with $\sum_H |l_{zH}|=1$ or $s^{-5}$ with $\sum_H |l_{zH}|=2$, while the helicity conserving amplitudes were known to scale as $s^{-4}$. The interference between amplitudes with different helicity changes offers a new mechanism to explain the spin-spin correlation and oscillation around the scaling value in $pp$ scattering~\cite{dutta04}. This can also lead to the deviation from the traditional constituent counting rule for other exclusive processes, such as photoproduction of charged pions that is discussed in this paper.

Therefore, a detailed investigation of the scaling behavior may enable a 
test of the generalized counting rule, though a more rigorous test should 
come from the polarization measurements, which would allow the 
separation of amplitudes with different helicity changes.

The deviation from the constituent counting rule for exclusive processes may also be due to the breakdown of the locality of quark-hadron duality~\cite{zhao03}. Quark-hadron duality is an empirical property of the data discovered by Bloom and Gilman before the advent of QCD~\cite{bloom}. The production of resonances at lower energies and momentum transfers averages smoothly around the scaling curve measured at large momentum transfers. The sum over resonances can be related to the scaling behavior as a result of destructive interference. This is rather local at high energy due to the high density of the overlapping resonances, which is called the locality of quark-hadron duality. But the local degeneracy may not be reached at energies of a few GeV, which leads to the restricted locality of quark-hadron duality. The restricted locality may cause oscillations around the scaling value above the resonance region when different partial waves are not canceled locally.

\begin{figure}[htpb]
\centerline{\includegraphics*[bb=25 32 296 150,width=8.0cm]{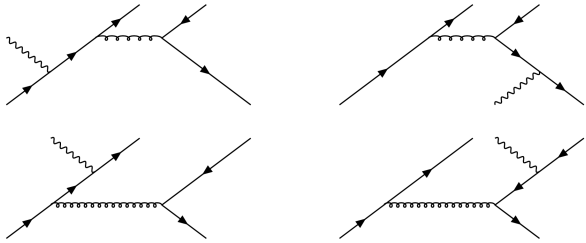}}
\caption{One-hard-gluon-exchange Feynman diagrams for the parton-level subprocess $\gamma q \rightarrow M q$ in the single meson photoproduction $\gamma N \rightarrow M N$.}
\label{ratio_th}
\end{figure}
The exclusive charged pion ratio of $\frac{d\sigma/dt(\gamma n \rightarrow \pi^- p)}{d\sigma/dt(\gamma p \rightarrow \pi^+ n)}$ can be estimated based on the one-hard-gluon-exchange Feynman diagrams~\cite{huang1,huang2}. In Huang {\it et al.}'s approach~\cite{huang1}, the helicity amplitude for the photoproduction of a meson was assumed to factorize into the parton-level subprocess amplitude and the nucleon form factors. Evaluating the four Feynman diagrams in Figure~\ref{ratio_th} gives the parton-level subprocess amplitude for pseudoscalar meson photoproduction.
Due to isospin invariance, the form factors are divided out and the exclusive charged pion ratio takes on a simple form,
\begin{equation}
\frac{ d\sigma/dt(\gamma n \rightarrow \pi^-p) }{d\sigma/dt(\gamma p \rightarrow \pi^+n) } \simeq  \left( \frac{ue_d+se_u}{ue_u+se_d} \right)^2,
\label{eq:ratio}
\end{equation}
 where $u$ and $s$ are the Mandelstam variables and $e_q$ denotes the charge of quark $q$.
 
\section{Experiment} 
 
To study the transition from nucleon-meson degrees of freedom to quark-gluon degrees of freedom, it is essential to investigate the GeV region where the transition appears to occur. While there were some measurements at SLAC for the $\gamma p \rightarrow \pi^+ n$ process at photon energies of 4, 5 and 7.5 GeV~\cite{anderson76}, which exhibit a global scaling behavior expected by the constituent counting rule, there are no data beyond 2 GeV for   
$\gamma n \rightarrow \pi^- p$~\cite{world_data,besch,world_ref}. This experiment, JLab experiment E94-104~\cite{E94-104} was proposed to measure the cross section for charged pion photoproduction ${d\sigma/dt(\gamma n \rightarrow \pi^- p)}$ and ${d\sigma/dt(\gamma p \rightarrow \pi^+ n)}$ from 1.1 to 5.5 GeV. In addition, the differential cross section ratio for charged pion photoproduction can be formed and compared to theoretical predictions. 

\begin{figure*}[htbp]
\vspace{-0.1in}
\centerline{\includegraphics[scale=0.5]{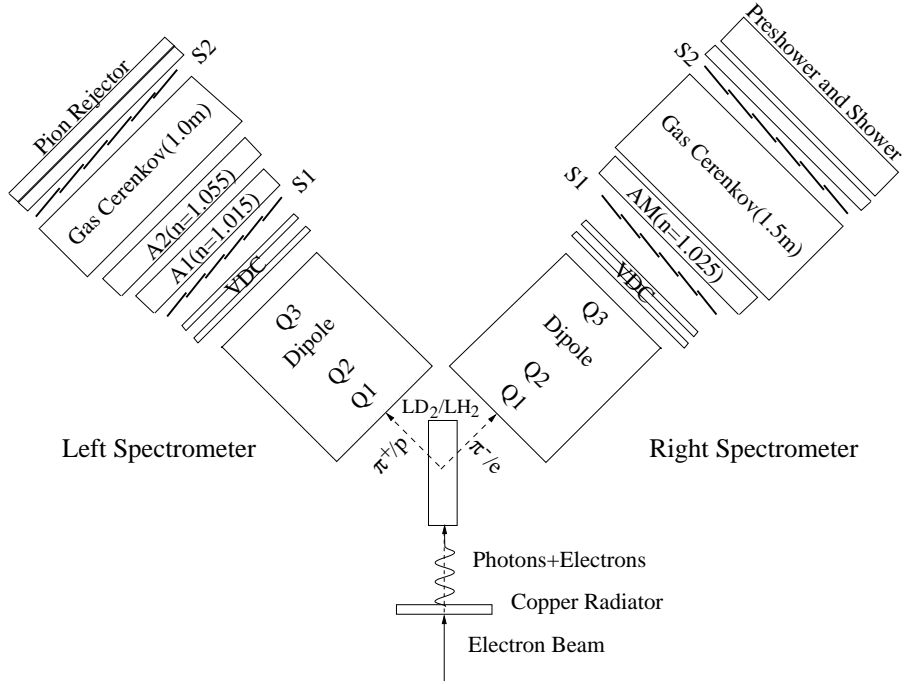}}
\caption{Schematic view of the experimental setup for E94-104.}
\label{fig:detector}
\end{figure*}

The experiment E94-104 was carried out in Jlab~\cite{leemann} Hall A~\cite{NIM}, using the continuous electron beam at currents around 30 $\mu$A. The core of the Hall A equipment is a pair of nearly identical 4 GeV/c spectrometers capable of determining the momentum and angles of charged particles with high resolution. The schematic view of the setup for this experiment is shown in Figure~\ref{fig:detector}. The real bremsstrahlung photons were generated by the electrons impinging on a copper radiator, located 72.6 cm upstream from the target. The foil with a thickness of 6.12\% radiation length was used for the production data of E94-104. A liquid hydrogen target (LH$_2$) was used as the proton target, while a liquid deuterium target (LD$_2$) was used as an effective neutron target.  The outgoing pions and protons were detected by the two high resolution spectrometers (HRS) in Hall A. The vertical drift chambers (VDCs) recorded its track and the scintillator planes (S1/S2) provided timing information and generated triggers.  Aerogel \v{C}erenkov detectors (A1/A2/AM) provided particle identification for positively charged particles, mainly pions and protons. Gas \v{C}erenkov, preshower/shower detector and pion rejector were used to discriminate negatively charged particles, mainly electrons and pions. The left spectrometer was optimized to detect positively charged particles while the right one was optimized to detect negatively charged particles. However, both spectrometers had to contain detectors to identify both negatively and positively charged particles, since there were a few reversed polarity kinematics. Based on two-body kinematics, the incident photon energies were reconstructed from the final states, i.e. the momentum and angle of the $\pi^+$ in the singles measurements for the $\gamma p \rightarrow \pi^+ n$ process, or momenta and angles of the $\pi^-$ and $p$ in the coincidence measurements for the $\gamma n \rightarrow \pi^- p$ process. 

The coincidence kinematics for the $\gamma n \rightarrow \pi^- p$ process are listed in Table~\ref{tab:kin_coin}. Normally, the negatively charged pions are detected by the spectrometer to the right of the beam line (viewed along the beam direction), and the protons detected by the left spectrometer. But a few kinematics require reversing the polarities of the spectrometers, because the maximum momentum of the right spectrometer is only 3.16 GeV/c, while that of the left spectrometer is 4 GeV/c. The singles kinematics for the $\gamma p \rightarrow \pi^+ n$ process are listed in Table~\ref{tab:kin_sing}. The positively charged pions were detected by the left spectrometer with positive polarity. The beam energies tabulated in Table~\ref{tab:kin_coin} and Table~\ref{tab:kin_sing} were the nominal values used to set the spectrometers, which may deviate from the measured ones by several MeV. The spectrometer momentum and angle settings were calculated by using a photon energy close to the beam energy, i.e. $E_e-75$ (MeV), where the multiple pion production processes were suppressed.

\begin{table*} 
\caption{Spectrometer settings for coincidence kinematics. $E_e$ is the electron beam energy, $\theta_{c.m.}$ the pion center-of-mass angle, $P_L$ ($P_R$) the central momentum for left (right) spectrometer with the sign indicating its polarity, and  
$\theta_L$ ($\theta_R$) is the central scattering angle for left (right) spectrometer. The Mandelstam variables $\sqrt{s}$ and $-t$ are in the last two columns.} 
\begin{center}
\begin{tabular}{|cccccccc|} 
\hline \hline 
$E_e$ & $\theta_{c.m.}$ & $P_L$ & $\theta_L$  & $P_R$ & $\theta_R$ & $\sqrt{s}$ & $-t$  \\  
(GeV)  &  ($^\circ$) & (GeV/c) & ($^\circ$) & (GeV/c) & ($^\circ$) & (GeV) & (GeV/c)$^2$  \\ 
  \hline \hline 
 
1.173 &   50.0 &   +0.521 &   60.17 &    -0.953 &   28.33 &1.71&0.253 \\ 
1.173 &   70.0 &   +0.727 &   49.72 &    -0.838 &   41.45 &1.71&0.467 \\ 
1.173 &   90.0 &   +0.923 &   39.75 &    -0.706 &   56.66 &1.71&0.709 \\ 
\hline 
  
1.721 &   50.0 &   +0.697 &   58.32 &    -1.433 &   24.46 & 1.99&    0.433 \\  
1.721 &   70.0 &   +0.989 &   47.39 &    -1.238 &   36.02 & 1.99&    0.798 \\  
1.721 &   90.0 &   +1.277 &   37.37 &    -1.015 &   49.73 & 1.99&    1.212 \\   
\hline

1.875 &   50.0 &   +0.742 &   57.79 &   -1.566 &   23.64 & 2.06&    0.484 \\  
1.875 &   90.0 &   +1.370 &   36.75 &   -1.099 &   48.21 & 2.06&    1.355 \\ 
\hline 
  
2.558 &   50.0 &   +0.913 &   55.67 &    -2.108 &   20.96 & 2.35&    0.696 \\  
2.558 &   70.0 &   +1.322 &   44.37 &    -1.794 &   31.02 & 2.35&    1.282 \\  
2.558 &   90.0 &   +1.740 &   34.45 &    -1.438 &   43.18 & 2.35&    1.948 \\  
2.558 &   50.0 &   -2.108 &   20.96 &    +0.913 &   55.67 & 2.35&    0.696 \\  
2.558 &   70.0 &   -1.794 &   31.02 &    +1.322 &   44.37 & 2.35&    1.282 \\  
2.558 &   90.0 &   -1.438 &   43.18 &    +1.740 &   34.45 & 2.35&    1.948 \\ 
\hline 
 
3.395 &   50.0 &    +1.113 &   53.21 &  -2.799 &   18.57 & 2.67&    0.971 \\  
3.395 &   70.0 &    +1.642 &   41.74 &  -2.363 &   27.56 & 2.67&    1.789 \\  
3.395 &   90.0 &    +2.195 &   32.01 &  -1.866 &   38.57 & 2.67&    2.718 \\  
3.395 &  100.0 &    +2.466 &   27.69 &  -1.614 &   45.24 & 2.67&    3.190 \\  
3.395 &  110.0 &    +2.725 &   23.65 &  -1.369 &   53.01 & 2.67&    3.648 \\ 
\hline 
 
4.232 &   50.0 &   -3.489 &   16.84 &   +1.300 &   51.04 & 2.95& 1.248 \\  
4.232 &   70.0 &   +1.949 &   39.51 &   -2.929 &   25.05 & 2.95&    2.299 \\  
4.232 &   90.0 &   +2.638 &   30.01 &   -2.291 &   35.18 & 2.95&    3.494 \\  
\hline 
 
5.618 &   70.0 &   -3.863 &   22.08 &   +2.442 &   36.48 & 3.36&    3.148 \\  
5.618 &   90.0 &   +3.359 &   27.38 &   -2.990 &   31.11 & 3.36&    4.785 \\  
\hline\hline 
 \end{tabular}
\end{center}  
\label{tab:kin_coin} 
\end{table*} 
\begin{table*} 
\caption{Spectrometer settings for singles kinematics. $E_e$ is the electron beam energy, $\theta_{c.m.}$ the pion center-of-mass angle, $P_L$ the central momentum for left spectrometer with the sign indicating its polarity, and  
$\theta_L$ is the central scattering angle for left spectrometer. The Mandelstam variables $\sqrt{s}$ and $-t$ are in the last two columns.} 
\label{tab:kin_sing} 
\begin{center}
\begin{tabular}{|cccccc|} 
\hline \hline 
$E_e$ & $\theta_{c.m.}$ & $P_L$ & $\theta_L$  &$\sqrt{s}$ & $-t$  \\  
(GeV)  &  ($^\circ$) & (GeV/c) & ($^\circ$) & (GeV) & (GeV/c)$^2$  \\ 
 
  \hline \hline 
  
1.173 &   70.0 &   +0.838 &   41.45 &   1.71&   0.467 \\  
1.173 &   90.0 &   +0.706 &   56.66 &  1.71&   0.709 \\  
\hline 
1.721 &   50.0 &  +1.433 &   24.46 &    1.99&    0.433 \\  
1.721 &   70.0 &  +1.238 &   36.02 &    1.99&    0.798 \\  
1.721 &   90.0 &  +1.015 &   49.73 &    1.99&     1.212 \\ 
\hline 
1.875 &   50.0 &   +1.566 &   23.64 &   2.06&    0.484 \\  
1.875 &   90.0 &   +1.099 &   48.21 &   2.06&     1.355 \\ 
\hline 
2.558 &   50.0 &   +2.108 &   20.96 &   2.35&    0.696 \\  
2.558 &   70.0 &   +1.794 &   31.02 &   2.35&    1.282 \\  
2.558 &   90.0 &   +1.438 &   43.18 &   2.35&     1.948 \\  
\hline 
3.395 &   50.0 &   +2.799 &   18.57 &   2.67&     0.971 \\  
3.395 &   70.0 &   +2.363 &   27.56 &   2.67&     1.789 \\  
3.395 &   90.0 &   +1.866 &   38.57 &   2.67&     2.718 \\  
3.395 &  100.0 &   +1.614 &   45.24 &   2.67&     3.190 \\  
3.395 &  110.0 &   +1.369 &   53.01 &   2.67&     3.648 \\ 
\hline 
4.232 &   70.0 &   +2.929 &   25.05 &   2.95&     2.299 \\ 
4.232 &   90.0 &   +2.291 &   35.18 &   2.95&     3.494 \\  
4.232 &  100.0 &   +1.967 &   41.36 &   2.95&     4.101 \\  
  
\hline 
5.618 &   90.0 &   +2.990 &   31.11 &   3.36&     4.785 \\  
5.618 &  100.0 &   +2.547 &   36.69 &   3.36&     5.615 \\  
\hline\hline 
\end{tabular} 
\end{center}
\end{table*}  
 
\section{Data analysis} 
\subsection{Overview}
The raw data from the data acquisition (DAQ) system were replayed or decoded by an event processing program, ESPACE (Event Scanning Program for Hall A Collaboration Experiments) using CERNLIB packages. The outputs were histograms and ntuples of physical variables in the HBOOK format. The yield from the data was obtained by applying cuts on certain variables in the ntuples, such as trigger type, particle type, spectrometer acceptance and reconstructed photon energy. Next, the yield was normalized by beam charge and computer deadtime. To extract the differential cross sections, simulations were carried out by using the modified MCEEP (\underline {M}onte \underline{C}arlo for $(\underline{e},\underline{e'}\underline{p})$) program~\cite{MCEEP} written for JLab Hall A. 
The raw differential cross section $(\frac{d\sigma}{dt})_{\rm data}$ was extracted by comparing the background subtracted yield from the data ($ Y_{\rm data}$) with the yield from the Monte Carlo simulation ($Y_{\rm mc}$): 
\begin{equation} 
(\frac{d\sigma}{dt})_{\rm data}= (\frac{d\sigma}{dt})_{\rm mc}* \frac{Y_{\rm data}}{ Y_{\rm mc}}. 
\label{compare}
\end{equation} 
Finally to extract the physical differential cross section, corrections such as the nuclear transparency of deuterium for the final state interaction, the detection efficiency and nuclear absorption in the detection materials were applied to the raw differential cross section. 

\subsection{Acceptance Analysis}
The R-function is defined to be the minimal distance to the acceptance boundary in terms of several two-dimensional polygons. It helps to select events in the central region of the spectrometer acceptance in a systematic and efficient way, where the optics matrix elements are well tuned. 
This method was originally developed in the E89-044 data analysis~\cite{TN01055}. The version~\cite{chai_th} refined in the E91-011 data analysis was used to analyze the E94-104 data and will be discussed below.

The R-function is generated for each event in both data analysis and simulation to optimize the cuts on different acceptance variables, i.e.  $\theta_{tg}$, $\phi_{tg}$, $y_{tg}$ and $\delta$. The $\theta_{tg}$ is the deviation of the out-of-plane angle from the spectrometer central setting of zero. The $\phi_{tg}$ is the deviation of the in-plane angle from the central setting of scattering angle. The $y_{tg}$ is the reaction vertex position in the target along the direction perpendicular to the spectrometers. The $\delta$ is the relative deviation from the central momentum setting. 
 
Six two-dimensional boundaries are defined for each spectrometer, out of any two combinations of the four acceptance variables, $\theta_{tg}$, $\phi_{tg}$, $y_{tg}$ and $\delta$. Each boundary is a polygon defined in a two-dimension plot of the data. For each event, the magnitude of the distance to the boundary is normalized based on the maximal length. The sign of the distance is positive for the events inside the polygon. The R-function for a single spectrometer, useful for singles $\gamma p \rightarrow \pi^+ n$ data, is defined to be the minimal distance to the six boundaries, while that for two spectrometers, useful for coincidence $\gamma n \rightarrow \pi^- p$ data, is defined by twelve two-dimension boundaries.

\subsection{Particle Identification Analysis}

For the detection of positively charged particles, the protons need to be selected for the coincidence $\gamma n \rightarrow \pi^- p$ process and the pions need to be selected for the singles $\gamma p \rightarrow \pi^+ n$ process. The aerogel \v{C}erenkov detectors A1/A2/AM (see Figure~\ref{fig:detector}) were utilized to identify protons and pions. The A1 and A2 detectors were used for normal polarity data, while AM was used for reversed polarity data. The particle identification with aerogel detectors was consistent with other methods available at low momentum, for example measuring the Time-Of-Flight of the particle or the energy deposited in the scintillators. 

For the detection of negatively charged particles, the pions need to be identified from the electron background for the coincidence $\gamma n \rightarrow \pi^- p$ process. It was realized by using the combination of the gas \v{C}erenkov and shower-type detector,  i.e. the gas \v{C}erenkov and total shower (preshower/shower) detector for normal polarity kinematics, the gas \v{C}erenkov and pion rejector detector for reversed polarity kinematics. A one-dimensional cut was used for gas \v{C}erenkov detector, while  
a two-dimensional graphic cut was defined for the shower-type detector.

\subsection{Background Subtraction} 
The production data contained various kinds of background, such as those from the electroproduction process and those from the end caps of the target. Therefore each complete kinematics consisted of four different configurations for data taking, \\
(1) Radiator in, production target,  \\ 
(2) Radiator in, background target, \\ 
(3) Radiator out, production target, \\ 
(4) Radiator out, background target. \\ 
For the coincidence $\gamma n \rightarrow \pi^- p$ process, the LD$_2$ target was used as the production target and the LH$_2$ target for the background subtraction.
For the singles $\gamma p \rightarrow \pi^+ n$ process, the LH$_2$ target was used as the production target and the dummy target for the background subtraction. 
 
The backgrounds were subtracted from the coincidence $\pi^-$ or singles $\pi^+$ production yield according to 
\begin{eqnarray} 
\pi^-:(Y_{\rm in,LD_2}-Y_{\rm in,LH_2})-f(E_\gamma)(Y_{\rm out,LD_2}-Y_{\rm out,LH_2} )     \    \nonumber \\   
\pi^+:(Y_{\rm in,LH_2}-Y_{\rm in,Dum})-f(E_\gamma)(Y_{\rm out,LH_2}-Y_{\rm out,Dum} )  
\end{eqnarray} 
where the in or out represents using or removing the radiator during the data taking. The factor $f(E_\gamma)$ is less than unity due to the interaction between the radiator and the electron beam. For most kinematics, the yield without the radiator was about one third of that with the radiator. The yield without the production target, especially for coincidence cases, was much smaller than the production yield. 

\subsection{Monte Carlo Simulation} 
The Monte Carlo simulation was performed by using MCEEP~\cite{MCEEP}, a computer program designed for coincidence $(e,e'X)$ experiments in Hall A. The MCEEP program employs a uniform random sampling method to populate the experimental acceptance. An event is defined as one combination of variables that completely specifies the reaction in the laboratory. The cross section is considered as the weight of the event. It was modified for the coincidence $n(\gamma,\pi^-p)$ and singles $p(\gamma,\pi^+)n$ processes. The program was also modified to generate the R-function to define the acceptance cuts.  

The momentum distribution of the neutron inside the deuterium target was considered in calculating the kinematics and cross section. The bremsstrahlung photon energy was randomly generated and the bremsstrahlung photon yield spectrum was calculated using thin-radiator calculation based on reference~\cite{matthews73}, which was later corrected by the thick-radiator calculation~\cite{matthews73,meekins,matthews81}. The pion survival factor was also included in the cross section calculation based on the average flight length and was later corrected by muon contamination.

 The differential cross section $\frac{d\sigma}{dt}$ for $\pi^+$ photoproduction at fixed center-of-mass angle was assumed to be 
\begin{equation}
 \frac{d\sigma}{dt}=\frac{0.69 F_{survive}}{(1+{\rm cos}\theta_{c.m.})^{4}(1-{\rm cos}\theta_{c.m.})^{5}}\cdot(\frac{s_0}{s})^{7},
\end{equation}
where the angular distribution was fitted to SLAC data in the several GeV region~\cite{anderson76}. The factor of 0.69 and $s_0=10.263$ came from these SLAC data at 5 GeV. The $s^{-7}$ dependence was consistent with the constituent counting rule. For $\pi^-$ photoproduction, there was another constant to account for the deviation from $\pi^+$ photoproduction, but the overall normalization factor did not affect the final cross section extraction. The pion survival factor $F_{\rm survive}$ was calculated by
         $
        F_{\rm survive}={\rm exp}(-\frac{\rm L}{\gamma_{\pi}\tau_{\pi}\beta_{\pi}c}) \ ,
         $
with the average flight length $L$ of 25 m and pion mean lifetime of $\tau_{\pi}=2.60\times10^{-8}$ s. 

\begin{figure}[htbp] 
\centerline{\includegraphics*[bb=25 151 530 656,scale=0.5]{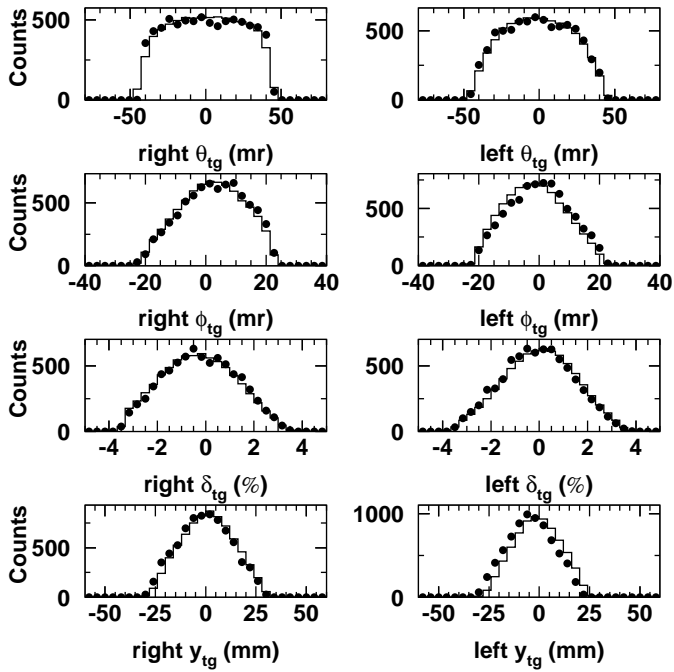}} 
\caption{Comparison of acceptance variables from left and right spectrometers between data and simulation for coincidence measurements. } 
\label{fig:comp_2} 
\end{figure}

\begin{figure}[htbp] 
\centerline{\includegraphics*[bb=20 150 530 670,scale=0.45]{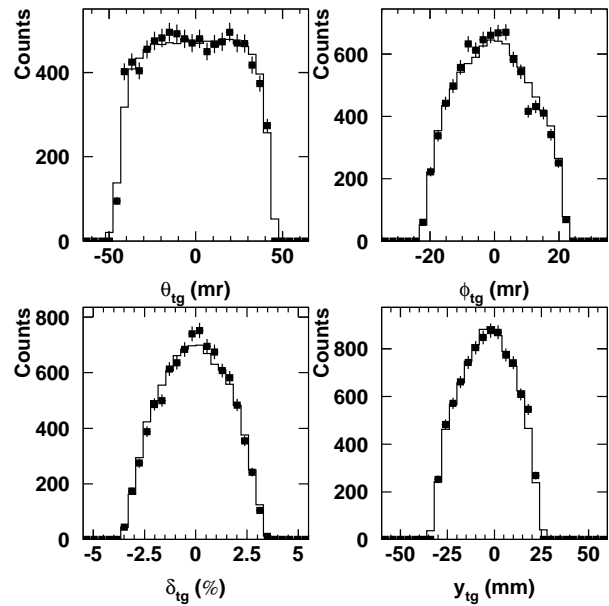}} 
\caption{Comparison of acceptance variables from left spectrometers between data and simulation for singles measurements. } 
\label{fig:sing_accpt} 
\end{figure}

\begin{figure}[htbp] 
\centerline{\includegraphics*[bb=12 394 290 670,width=4.5cm]{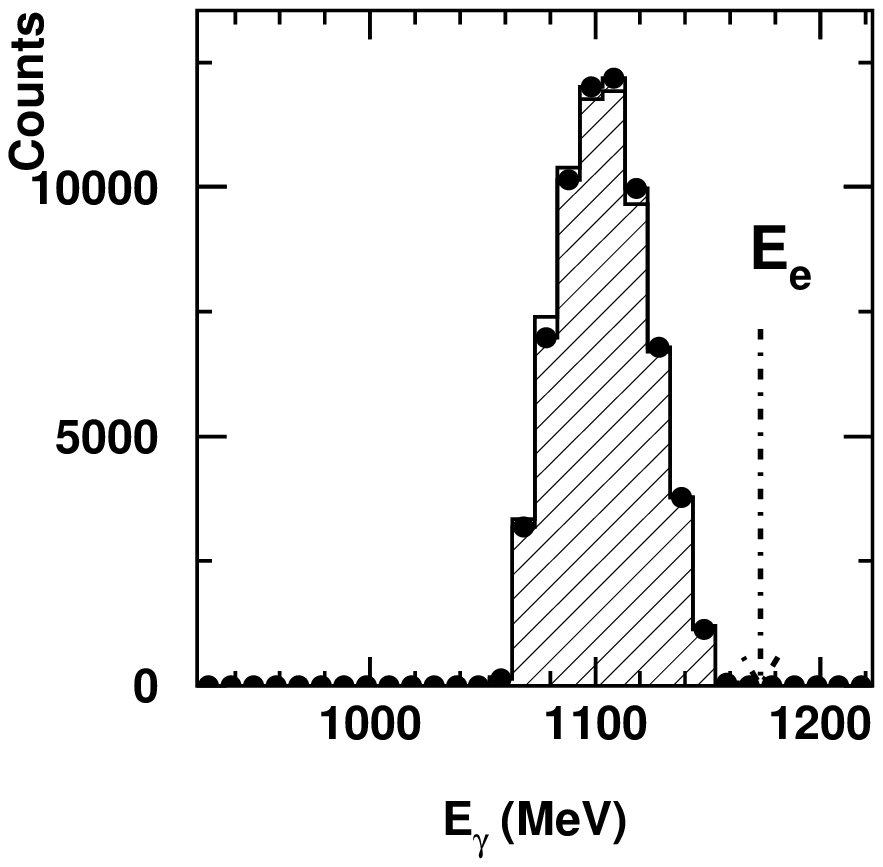} \includegraphics*[bb=12 394 290 670,width=4.5cm,height=4.5cm]{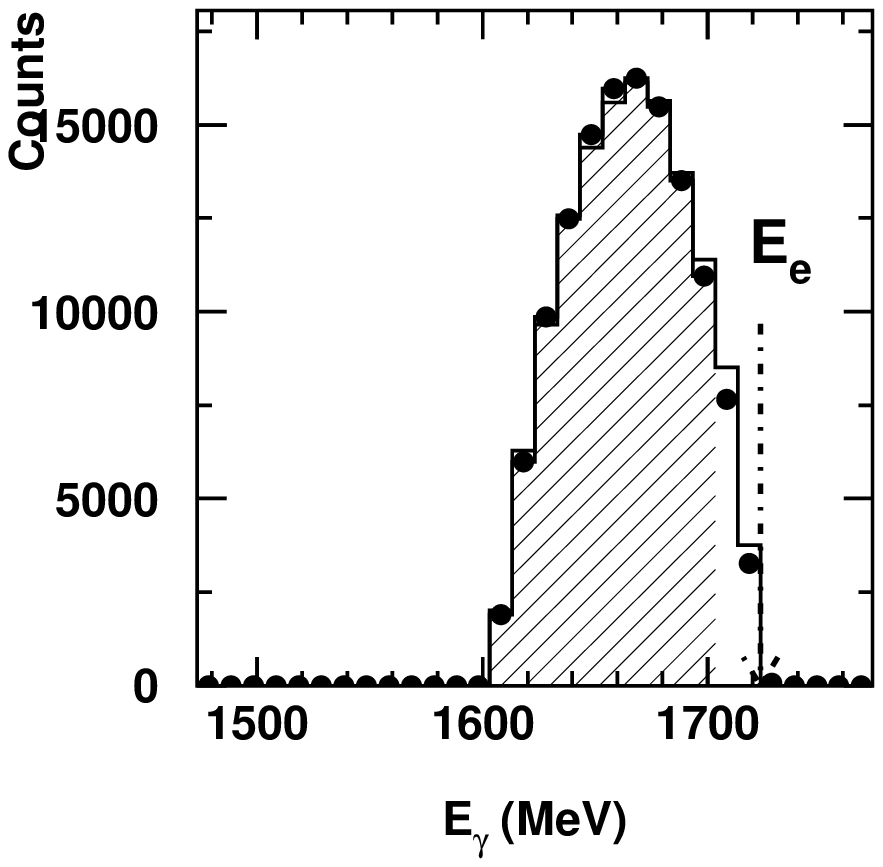}}
\vspace{-0.05in} \centerline{\vspace{-0.05in} \hspace{0.3in} (a) \hspace{1.5in} (b)}
\centerline{\includegraphics*[bb=12 394 290 670,width=4.5cm]{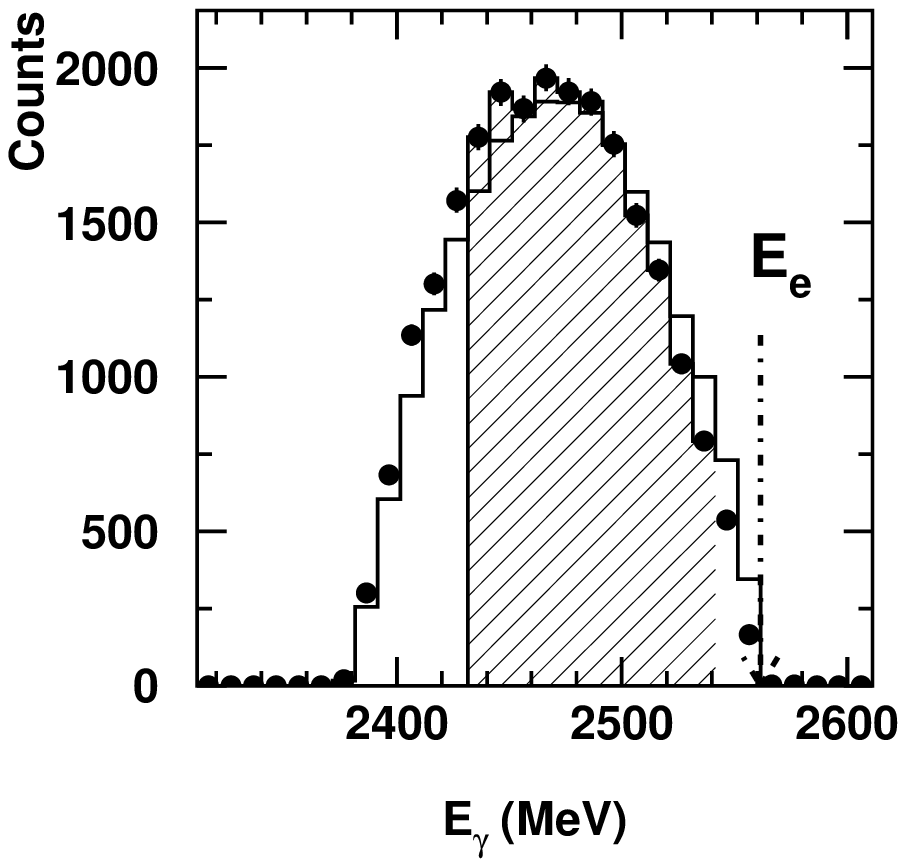} \includegraphics*[bb=12 394 290 670,width=4.5cm]{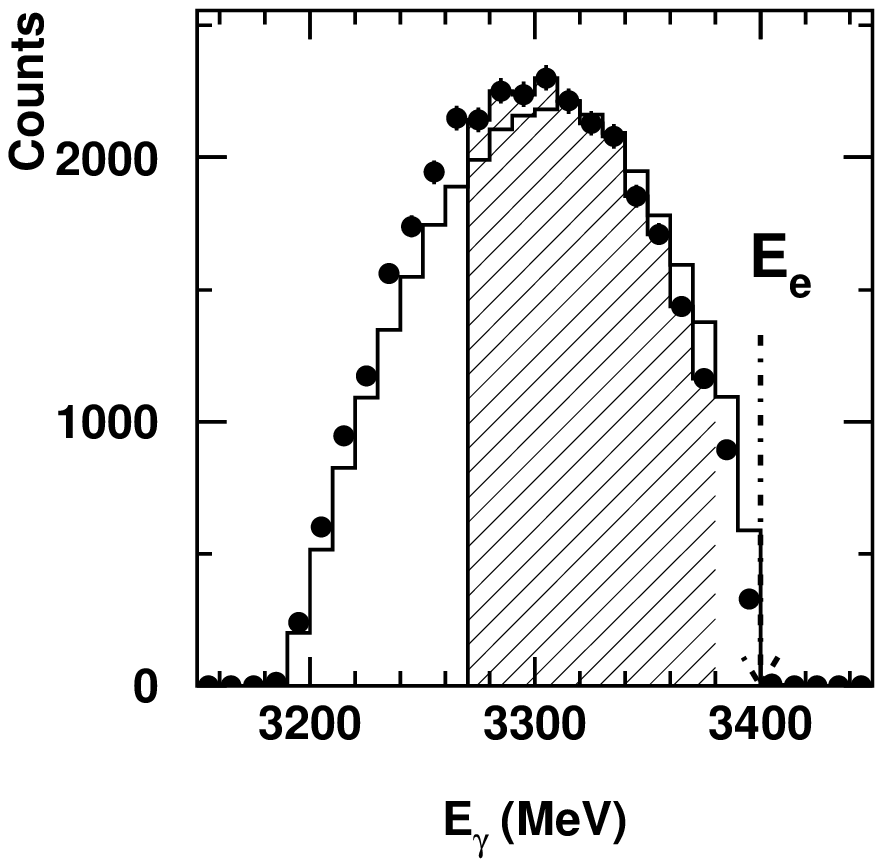}}
\vspace{-0.05in} \centerline{\vspace{-0.05in} \hspace{0.3in} (c) \hspace{1.5in} (d)}
\centerline{\includegraphics*[bb=12 394 290 670,width=4.5cm]{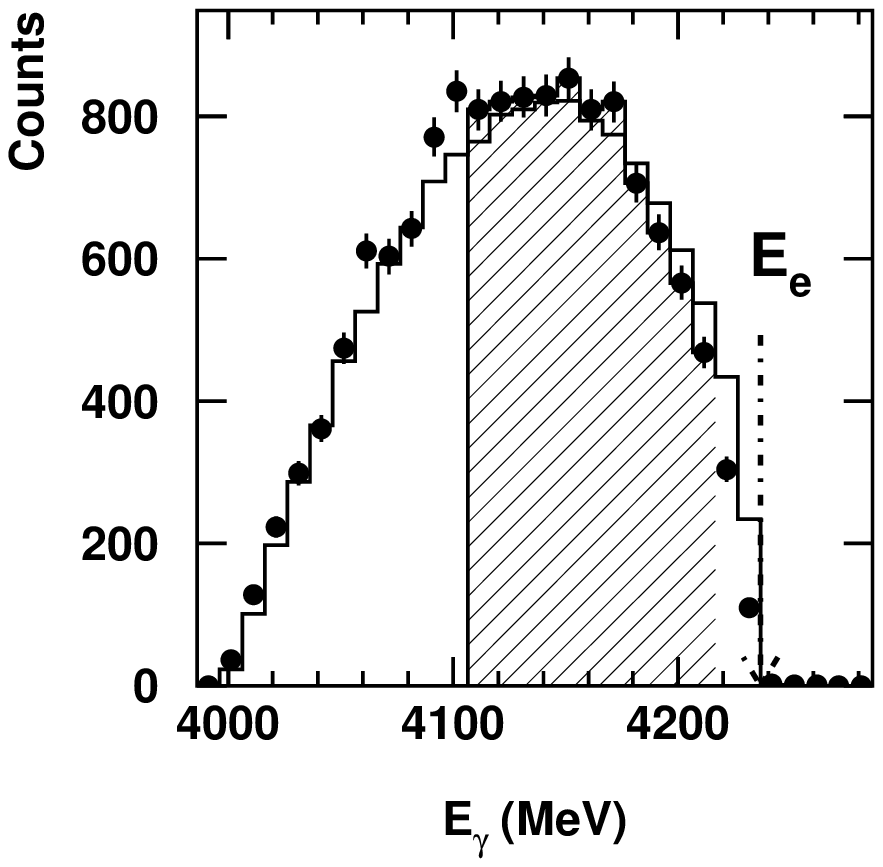} \includegraphics*[bb=12 394 290 670,width=4.5cm ]{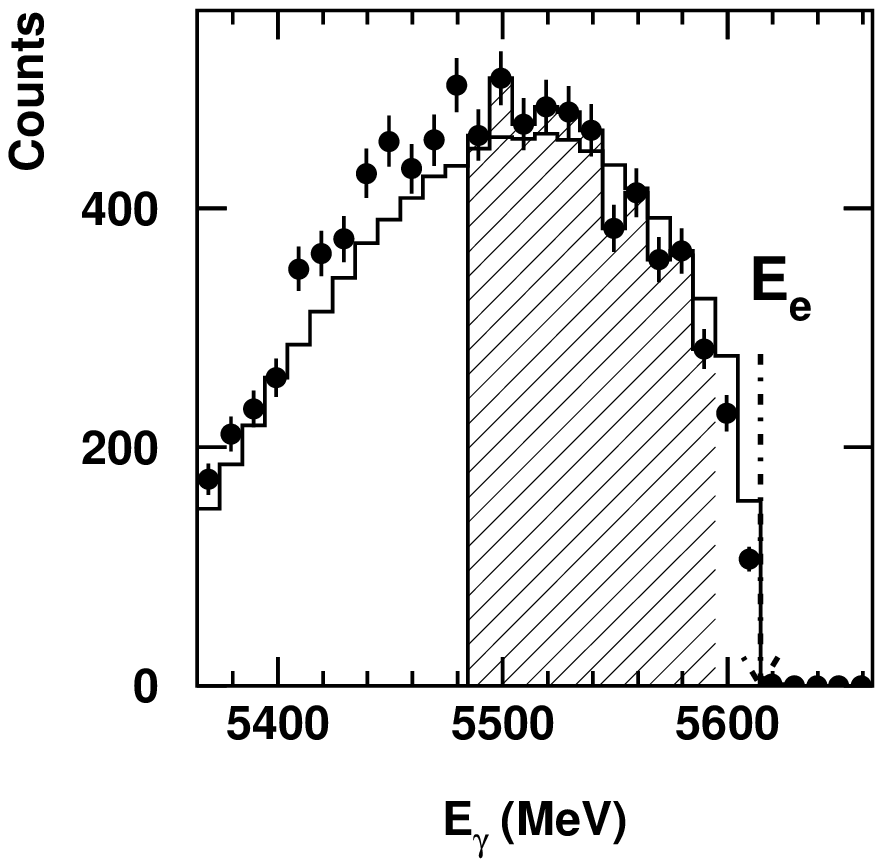}}
\vspace{-0.05in} \centerline{ \vspace{-0.05in} \hspace{0.3in} (e) \hspace{1.5in} (f)}
\caption{Comparison of reconstructed photon energy between data and simulation for coincidence measurements at $\theta_{c.m.}=90^\circ$. The results from data are plotted as symbols, while those from simulation are plotted as lines. The electron beam energies are 1173.3, 1723.4, 2561.5, 3400.0, 4236.4 and 5614.4 MeV. The comparison at beam energy 1876.9 MeV (not shown here) is very similar to that at 1723.4 MeV. The shaded events were chosen to extract the differential cross section.} 
\label{fig:comp_1} 
\end{figure}

\begin{figure}[htbp] 
\centerline{\includegraphics*[bb=31 394 300 670,width=4.4 cm ]{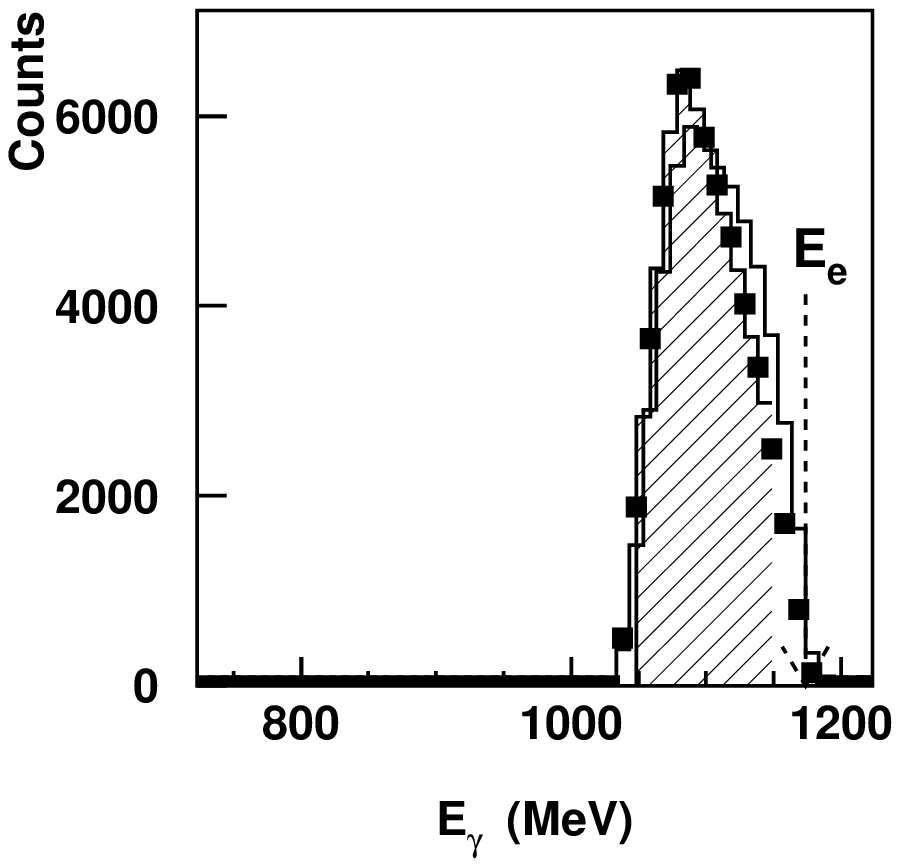} \includegraphics*[bb=31 394 300 670,width=4.4 cm ]{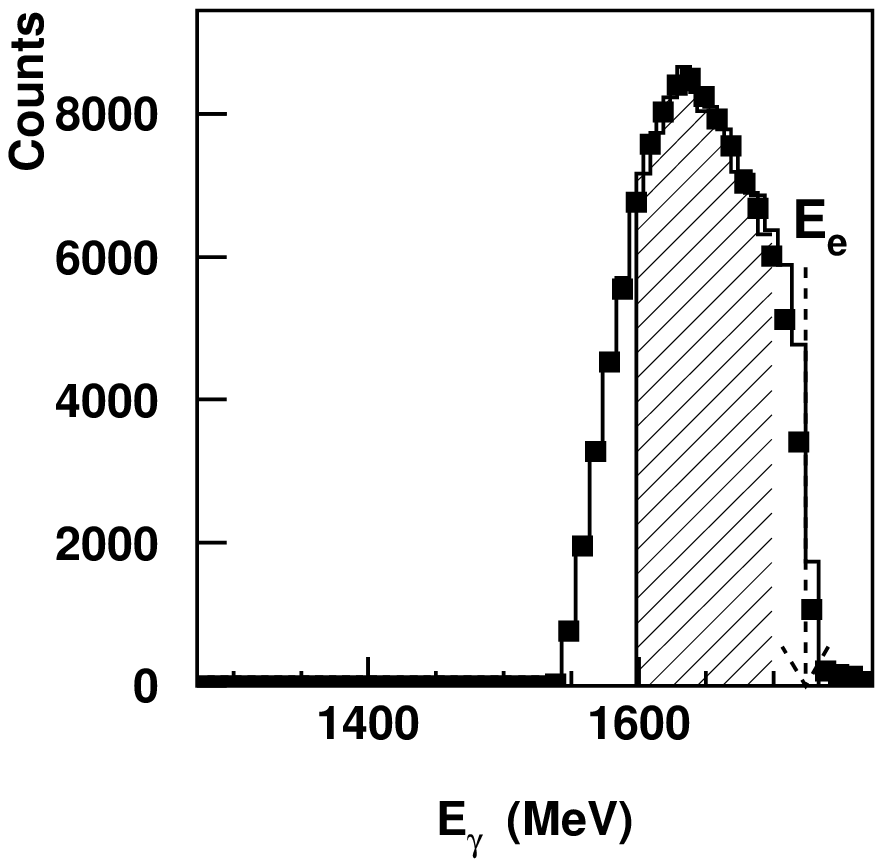}} 
\vspace{-0.05in} \centerline{\vspace{-0.05in} \hspace{0.0in} (a) \hspace{1.6in} (b)}
\centerline{\includegraphics*[bb=31 394 300 670,width=4.4 cm ]{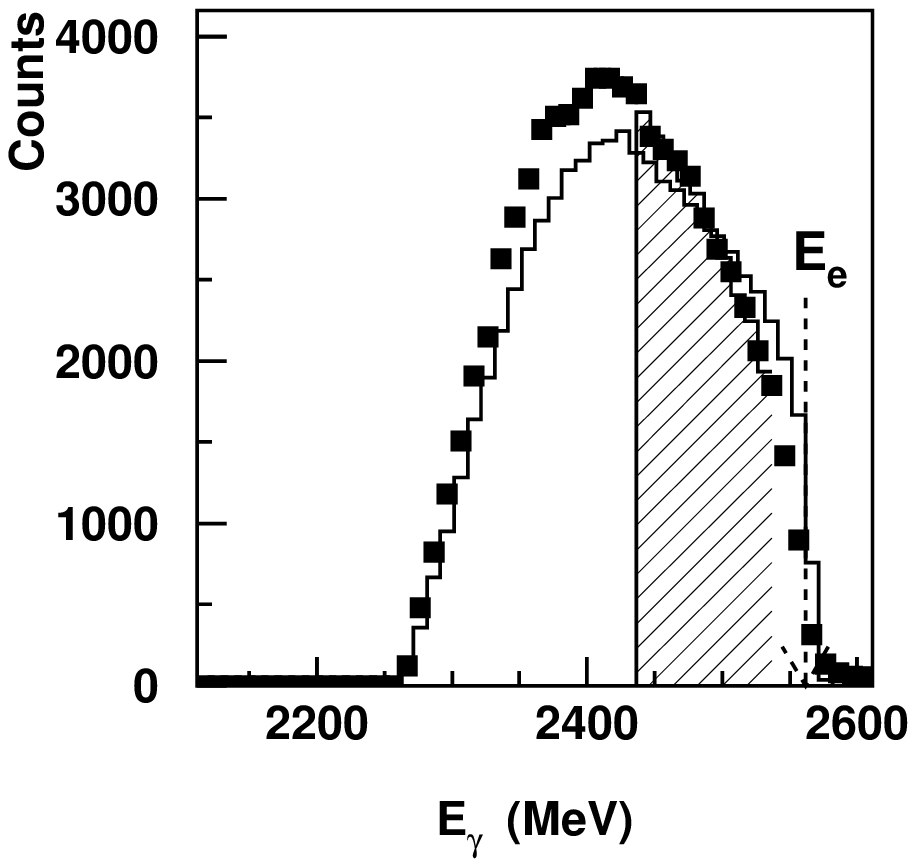} \includegraphics*[bb=31 394 300 670,width=4.4 cm ]{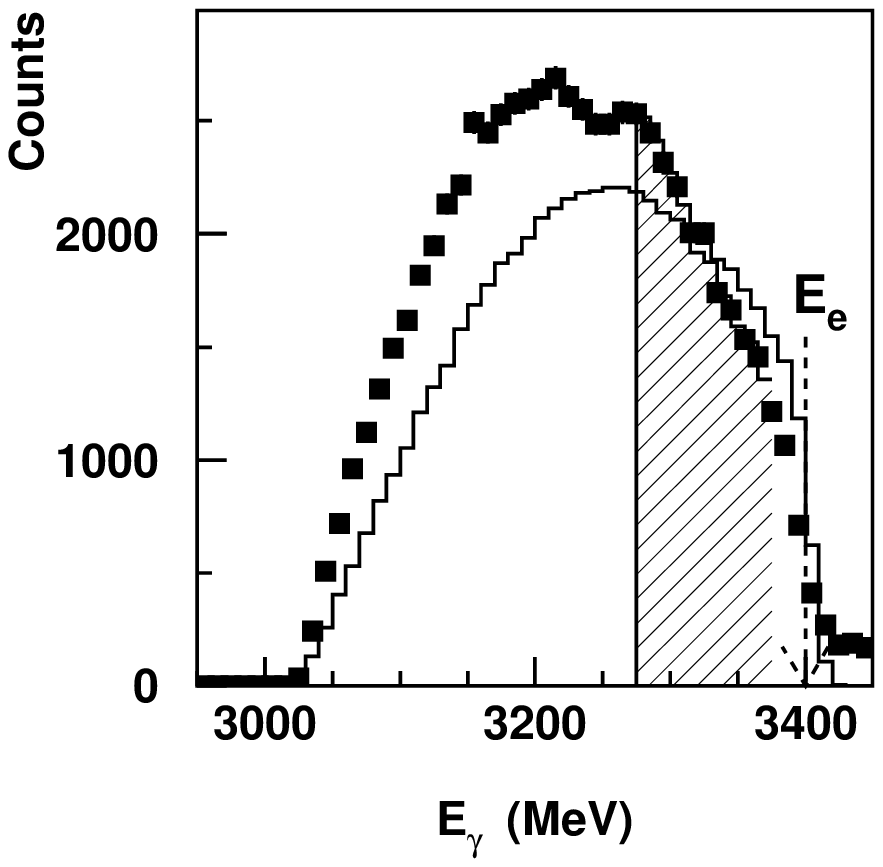}} 
\vspace{-0.05in} \centerline{\vspace{-0.05in}  \hspace{0.0in} (c) \hspace{1.6in} (d)}
\centerline{\includegraphics*[bb=31 394 300 670,width=4.4 cm ]{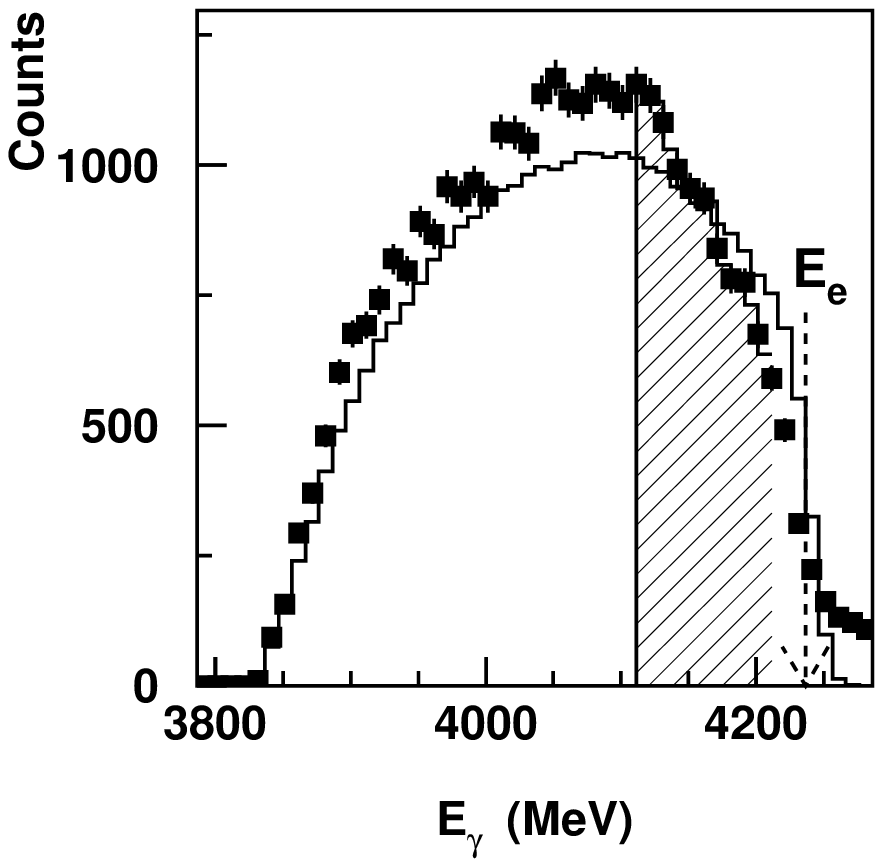} \includegraphics*[bb=31 394 300 670,width=4.4 cm ]{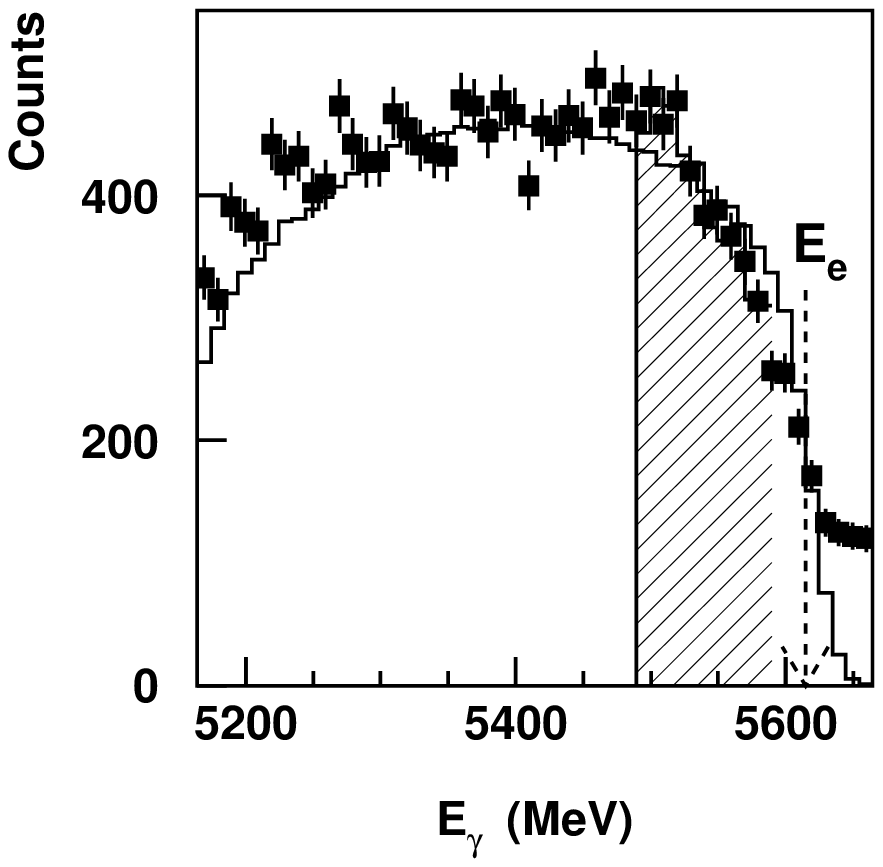}} 
\vspace{-0.05in} \centerline{\vspace{-0.05in} \hspace{0.0in} (e) \hspace{1.6in} (f)}
\caption{Comparison of reconstructed photon energy between data and simulation for singles measurements at $\theta_{c.m.}=90^\circ$. The results from data are plotted as symbols, while those from simulation are plotted as lines. The electron beam energies are 1173.3, 1723.4, 2561.5, 3400.0, 4236.4 and 5614.4 MeV. The comparison at beam energy 1876.9 MeV (not shown here) is very similar to that at 1723.4 MeV. The shaded events were chosen to extract the differential cross section.} 
\label{fig:sing_egamma} 
\end{figure}

\begin{figure}[htbp] 
\centerline{\includegraphics*[bb=35 150 530 656,scale=0.45]{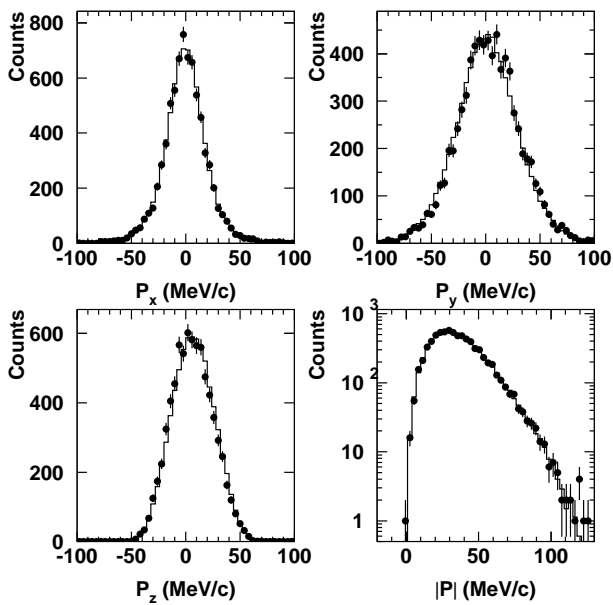}} 
\caption{Comparison of reconstructed momentum distribution of the neutron in the deuterium target between data and simulation for coincidence measurements. } 
\label{fig:comp_3} 
\end{figure}

\begin{figure}[htbp] 
\centerline{\includegraphics*[bb=31 400 530 670,scale=0.45]{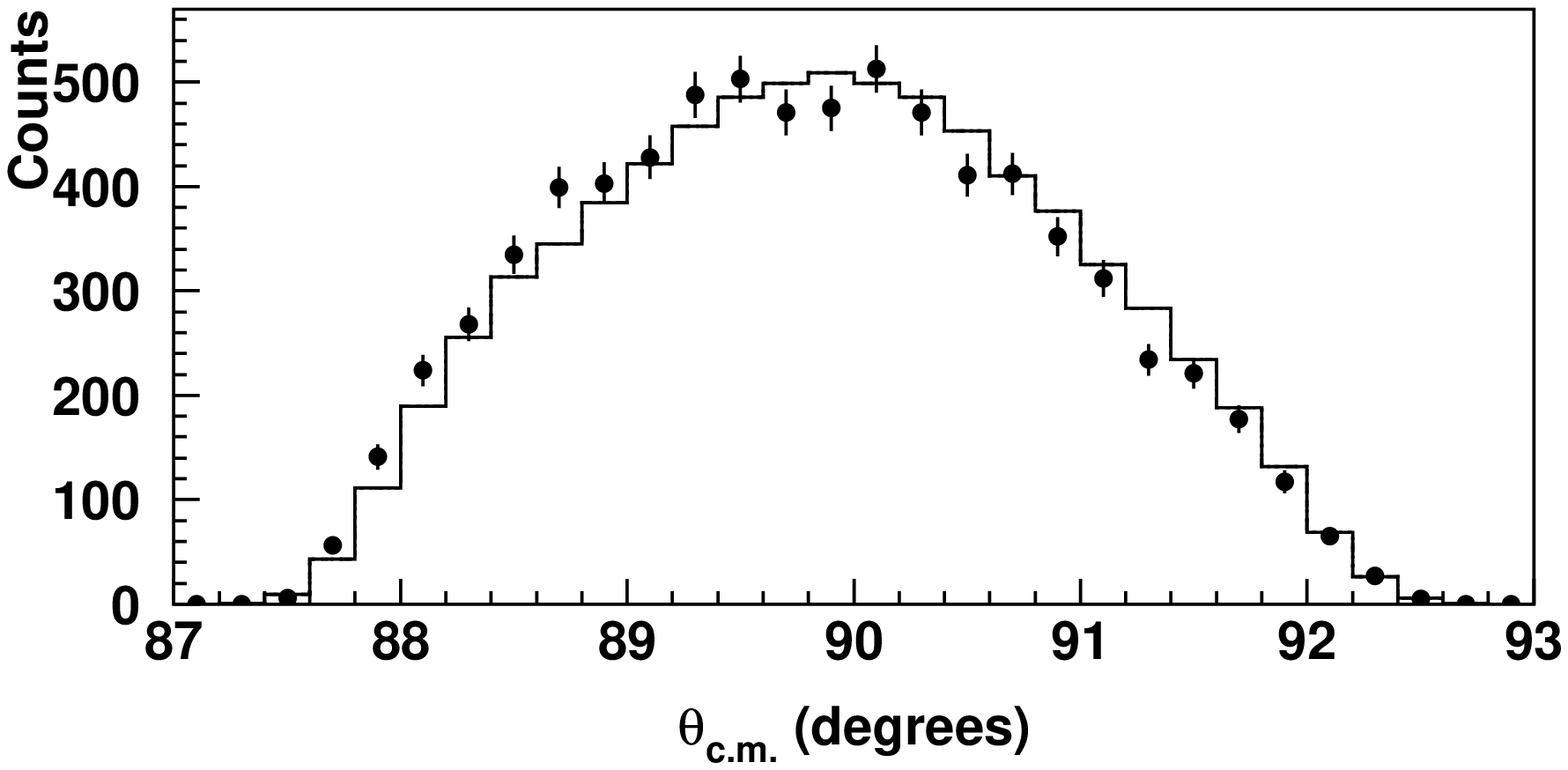}} 
\caption{Comparison of reconstructed center-of-mass angle between data and simulation for coincidence measurements. The nominal center-of-mass angle is 90 degrees.} 
\label{fig:coin_thetacm} 
\end{figure}

\begin{figure}[htbp] 
\centerline{\includegraphics*[bb=31 400 530 670,scale=0.45]{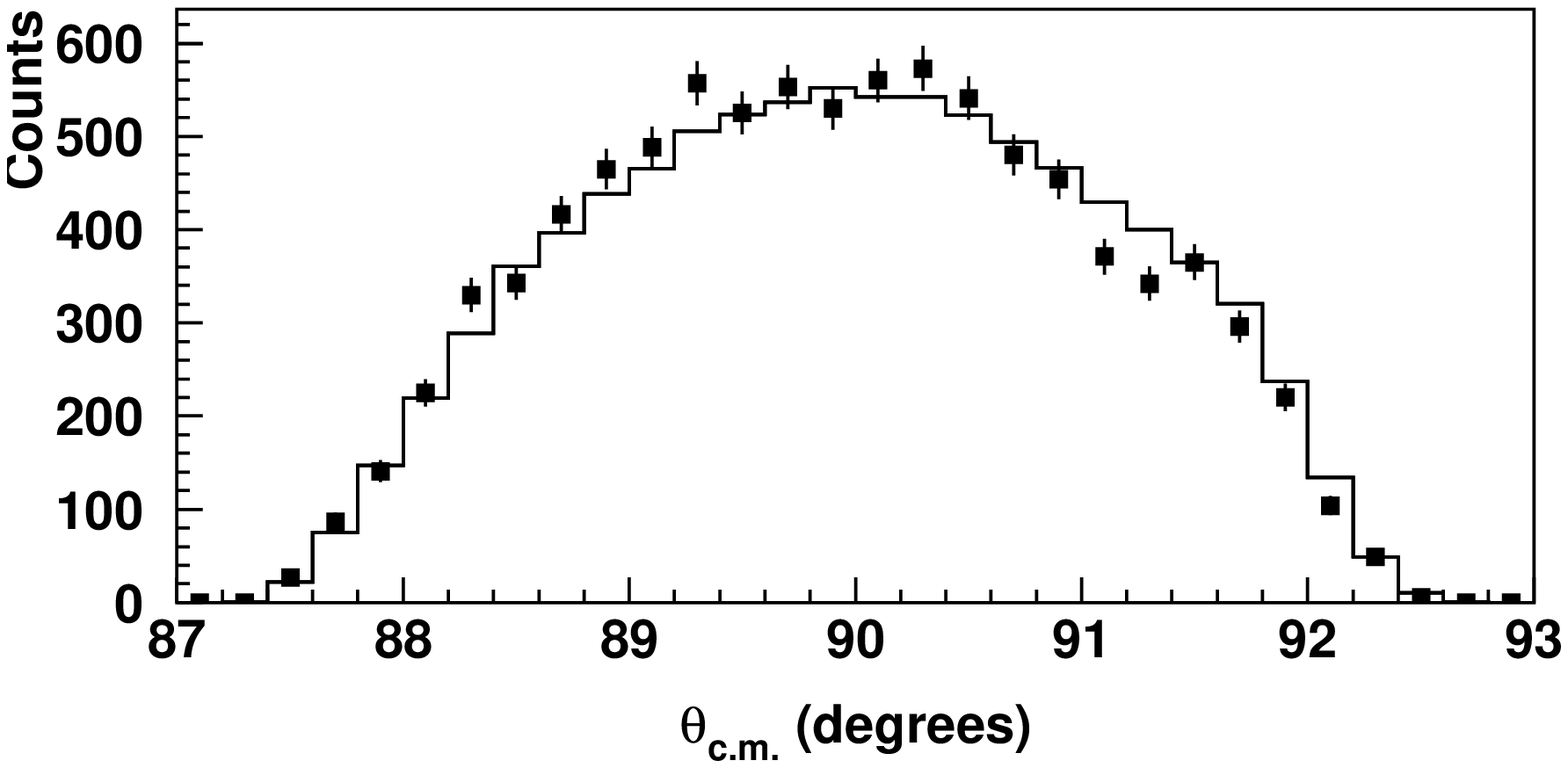}} 
\caption{Comparison of reconstructed center-of-mass angle between data and simulation for singles measurements. The nominal center-of-mass angle is 90 degrees.} 
\label{fig:sing_thetacm} 
\end{figure}

The distributions of different variables from the data were compared with those from the Monte Carlo simulation, including the acceptance variables, the reconstructed photon energy, the reconstructed initial momentum distribution of neutron inside the deuterium target, and the reconstructed center-of-mass angle, as shown in Figure~\ref{fig:comp_2} to Figure~\ref{fig:sing_thetacm}.
The results from data are symbolized by solid circles for coincidence measurements or by solid squares for singles measurements, while those from simulation are plotted as lines. Each simulated spectrum was multiplied by an arbitrary normalization factor for the comparison. This normalization factor is directly related to the cross section extraction, as shown in Equation~\ref{compare}. 
Except for the photon energy comparisons, only one typical kinematic setting is shown for other cases, i.e. the one at $E_e=4236$ MeV and $\theta_{c.m.}=90^\circ$ for the coincidence measurements, or at $E_e=4236$ MeV and $\theta_{c.m.}=90^\circ$ for the singles measurements.

 The overall agreement between data and simulation is good, but there is some discrepancy for some cases.  Since only reaction of interest was considered in the simulation, some discrepancy may be due to other physical reactions such as the multi-pion production. But the contribution from the multi-pion production is negligible with the cut on photon energy, as shown by the shaded area in  Figure~\ref{fig:comp_1} and Figure~\ref{fig:sing_egamma}. The uncertainties associated with various measurements such as the acceptance will affect the agreement too, which was included into the systematic uncertainties.


\subsection{Corrections and Uncertainties} 
Table~\ref{tab:correction} lists the systmatic uncertainties for kinematics at $\theta_{c.m.}=90^\circ$. The kinematics at other angles have similar breakdowns.  
 
\begin{table}[htbp] 
\caption{The systematic uncertainties for coincidence and singles measurements at $\theta_{c.m.}=90^\circ$. } 
\label{tab:correction} 
\begin{center}
\begin{tabular}{|c|c|c|}
\hline \hline
 Kinematics      & Coincidence (\%) & Singles (\%) \\
\hline \hline
Photon Yield      & 3.0 & 3.0\\
 \hline
$f(E_\gamma)$    & 1.0 & 1.0 \\  \hline
Target Density    &1.0 &1.0 \\  \hline
Tracking          & 3.0 & 2.0\\  \hline
Scintillator/Trigger & 1.3 & 1.0 \\ \hline
PID               & 2.2 & 3.0\\ \hline
Muon Contamination & 3.0 & 3.0\\ \hline
Transparency    &5.0 &-\\ \hline
Nuclear Absorption & 4.2 & 3.0\\ \hline
Trial Cross Section    & 2.2 & 2.2\\ \hline
Deadtime    & 2.0 & 2.0\\ \hline
Acceptance    & 3.0 & 3.0\\ \hline
Energy Loss    & 2.0 & 2.0\\ \hline
Random Coincidence    &1.0 &-\\ \hline
Momentum Distribution    &1.0 &-\\ \hline
Beam Charge    & 1.0 & 1.0\\ \hline
\hline 
Total    &10.1 &8.1 \\ \hline
\hline
\end{tabular}
\end{center}
\end{table} 
 
The thin-radiator calculation embedded in the simulation to calculate the photon flux was corrected for the radiator thickness, by using a thick-radiator code~\cite{meekins} based on references~\cite{matthews73, matthews81}, which considers the energy loss of electrons in the radiator and is expected to be accurate to a level of 3\% for the energies and radiator thickness used in this experiment. For the 6.12\% copper radiator, the thin-radiator calculation overestimates the yield by 11\% to 20\%.  By using this code~\cite{meekins}, the $f(E_\gamma)$ can be also calculated, which will affect the subtraction of electron induced background. 

The decrease of target density due to beam induced local boiling was corrected.
According to a comparison of the normalized yields with different beam currents, this correction was proportional to the average beam current and 
 can be parameterized as $0.072*I$ (\%) for LD$_2$ target and $0.048*I$ (\%) for the LH$_2$ target with $I$ the beam current in the unit of $\mu$A. The uncertainty was on the level of 1\%.

The single wire efficiency of the VDCs, with samples defined by two neighboring wires, was very close to 100\%. But multi-track events, dominated by two-track events, may still cause inefficiency in tracking. The corrections were applied based on the ratios of multi-track to single-track events, which were less than 2\% for most kinematics. The additional inefficiency associated with the track reconstruction algorithm was included in the uncertainty of 3\%.

The scintillator/trigger efficiency was obtained by checking the trigger for those selected events with good signals in VDCs and PID detectors. Special data were take to measure the efficiency, which was averaged to be 98.8$\pm$1.1\% for  the right spectrometer and 98.8$\pm$0.7\% for the left spectrometer.

One need to identify protons and negative pions for coincidence $\pi^-$ photoproduction, and identify only positively charged pions for singles $\pi^+$ photoproduction. There is no correction on proton identification for coincidence $\pi^-$ photoproduction due to the very high $p/\pi^+$ ratio ( $> 100$ ). The uncertainty due to $\delta$-electrons (or knock-on electrons) was estimated to be 1\%.
 For pion selection from coincidence $\pi^-$ photoproduction and singles $\pi^+$ photoproduction, two PID detectors were used. The correction and uncertainty were estimated based on the performance study of each detector by defining good event samples based on the other one. The corrections depended on the particle momentum and the signal to noise ratio, and therefore varied by kinematics. 

The pion decay loss was considered as the survival factor in the simulation by using the pion flight length and pion lifetime. The survival factor ranges from 53\% to 89\% for different pion momentum. However, since some of the muons from pion decay may still fall into the acceptance and be misidentified as pions, the calculation above may underestimate the effective pion survival factor and should be corrected. Based on the estimation with the modified simulation program SIMC~\cite{SIMC}, the correction depended on the particle momentum and ranged from -7\% to -4\% approximately~\cite{arrington}.
The uncertainty was estimated to be 3\% by checking the dependence of the correction on the acceptance cuts. 
 
Nuclear effects must be considered to obtain the cross section for $\gamma n \rightarrow \pi^{-} p$ from the measurement of $d(\gamma,\pi^-p)p$. The cross section may be reduced due to the final state interactions with the spectator proton inside the deuteron. The nuclear transparency was defined to be the ratio of the reduced cross section to the raw cross section without any final state interactions.  The measured nuclear transparencies for $d(e,e'p)$ quasi-elastic scattering~\cite{garrow}, show little $Q^2$ dependence above $Q^2 \simeq 2 $ (GeV/c)$^2$) and agree well with a Glauber calculation~\cite{glauber}. The fitted value of 0.904$\pm$0.013 was used to deduce the nuclear transparency for the experiment E94-104, based on the Glauber formulation.
The nuclear transparency for  $d(\gamma,\pi^-p)p$ was scaled from that of $d(e,e'p)n$ by replacing the total $pn$ scattering cross section with the total $pp$ and $\pi^-p$ scattering cross sections. 
The systematic uncertainties in the nuclear transparency were estimated to be 5\% to account for the uncertainties in the nuclear transparency measurement for $d(e,e'p)n$  and those in the effective $pn$, $pp$ and  $\pi^-p$ scattering cross sections. 
  
The produced particles, pions and protons in the coincidence measurements and pions in the singles measurements, had to go through various materials in the target and spectrometers before being detected. The event loss in the material is called nuclear absorption here. The major sources of nuclear absorption for high energy protons are listed in Table~\ref{tab:absorption}. The absorption was calculated based on the thickness and effective absorption length of the material in the flight path of the produced particles. The effective absorption length $\bar{\lambda}$ was estimated from the nuclear collision length $\lambda_T$ and nuclear interaction length  $\lambda_I$~\cite{pdg} as $2 \lambda_T \lambda_I/(\lambda_T+\lambda_I)$ by assuming that half of the elastic and quasi-elastic scattering contribute to the absorption. Later, the nuclear absorption was adjusted due to different flight lengths in the target, and different effective absorption lengths for pions and protons at various momenta. The flight length can be calculated from the scattering angle and the geometry of the target. The energy dependence of the effective absorption length was obtained from the cross section data in Reference~\cite{pdg}. The uncertainty for each produced hadron was estimated to be 3\%. 
  
\begin{table}[htbp] 
\caption{Major nuclear absorption in the target and spectrometer for high energy protons.} 
\label{tab:absorption} 
\begin{center} 
\begin{tabular}{|l|ccc|c|} 
\hline \hline 
Material  & Thickness  & Density & $\bar{\lambda}$& Absorp. \\ 
          & (cm)       & (g/cm$^3$) & (g/cm$^2$)  & (\%) \\ 
\hline 
15cm LD$_2$ (19K) & 5.5  & 0.1670  & 49.8 &1.84 \\ 
15cm LH$_2$ (22K) & 5.5  & 0.0723  & 46.8 &0.85 \\ 
\hline 
air                  & 300.  & 1.21e-3 & 73.4 &0.49 \\ 
S1 (Polystyrene)     & 0.5  & 1.032   & 68.3 &0.76 \\ 
S2 (Polystyrene)     & 0.5  & 1.032   & 68.3 &0.76 \\ 
\hline 
A1 (Aerogel)         & 9.0  & 0.060   & 75.5 &0.72 \\ 
A2 (Aerogel)         & 5.0  & 0.220   & 75.5 &1.46 \\ 
\hline 
AM (Aerogel)         & 9.0  & 0.100   &75.5 &1.19 \\ 
Gas \v{C}erenkov (CO$_2$) &150. &1.98e-3 & 73.6 &0.40\\ 
\hline \hline 
\end{tabular}  
\end{center}  
\end{table} 

The final cross section, extracted by comparing data and simulation, depended on the angular distribution and energy dependence of the trial cross section as the input of the simulation. Instead of searching for the exact form of the angular distribution and energy dependence of the actual cross section, the cross section fitted to SLAC data at high energy~\cite{anderson76} was used and its deviation from the actual cross section was considered as the systematic uncertainty. The trial angular distribution in the simulation may not be the same as the real case. The resulting systematic uncertainty was estimated by checking the change of the final cross section from the flat angular distribution. The changes of the final results were very small due to the small acceptance of the Hall A spectrometers. In the comparison, the mean center-of-mass angles were determined from the data, which deviated from those determined from the simulation by $0.2^\circ$ at most. The systematic uncertainties due to the angular distribution were estimated to be 1\%, 2\% and 3\% for kinematics at $\theta_{c.m.}=90^\circ,100^\circ$, at $\theta_{c.m.}=70^\circ,110^\circ$, and at $\theta_{c.m.}=50^\circ$ respectively. The trial cross section used in the simulation had a $s^{-7}$ energy dependence, which was suggested by the SLAC data at high energy~\cite{anderson76} and was predicted by the constituent counting rule. The actual energy dependences for both coincidence and singles kinematics are shown in Section~\ref{results}. The data, especially at low energy, do not have the $s^{-7}$ energy dependence.  The resulting systematic uncertainty was assigned to be 2\%, by checking the change of the final cross section from the flat energy dependence. 

The computer deadtime was calculated by taking the ratio of missed triggers in the data stream from DAQ to the input triggers from scalers. It was less than 20\% for nearly all the data and was corrected run-by-run (not listed here). 
The uncertainty was estimated to be around 10\% of the correction.
The electronics deadtime was less than 0.5\% for the majority of E94-104 data based on measurements using test pulses. There were also other systematics uncertainties for example due to acceptance, energy loss, random coincidence subtraction and beam charge. 

\subsection{Reversed Polarity Data} 
\begin{figure}[htbp] 
\centerline{\includegraphics*[bb=12 394 290 670,width=4.5cm]{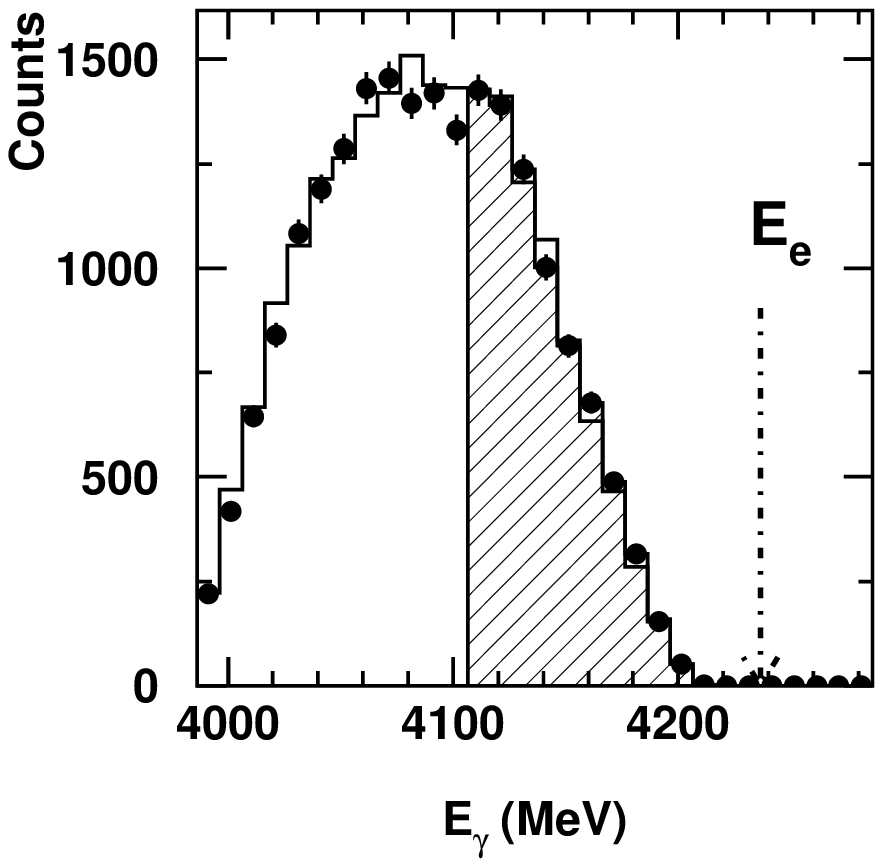} \includegraphics*[bb=12 394 290 670,width=4.5cm]{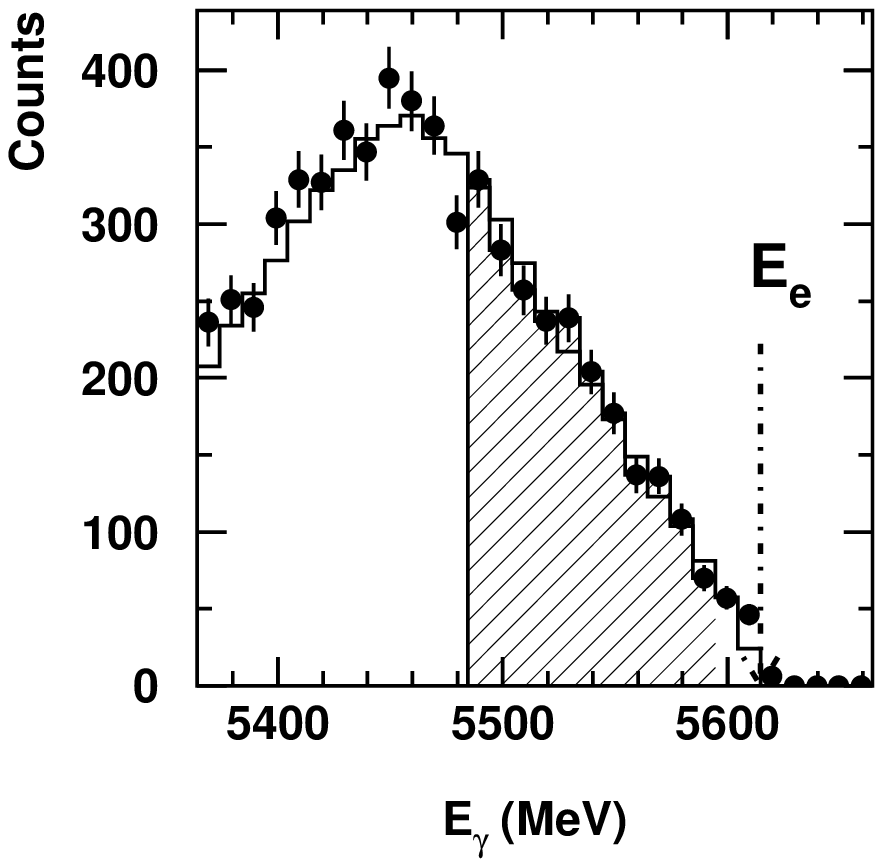}}
\vspace{-0.05in} \centerline{ \vspace{-0.05in} \hspace{0.1in} (a) \hspace{1.6in} (b)}
\caption{Comparison of reconstructed photon energy between data and simulation for coincidence $\pi^-$ photoproduction with reversed polarities. The plot on the left represents the kinematics at $E_e=4236.4$ MeV and $\theta_{c.m.}=50^\circ$, while the plot on the right represents the kinematics at $E_e=5614.4$ MeV and $\theta_{c.m.}=70^\circ$.} 
\label{fig:comp_rev} 
\end{figure}

The polarities of the spectrometers were reversed for a few kinematics during the experiment.  The particle identification was optimized for data acquisition with normal polarities. The aerogel detector (AM) in the right spectrometer was not as good as the combination of two aerogel detectors (A1 and A2) in the left spectrometer in identifying protons. But since the proton signals are very clean with very low pion background, this hardly affected the results. The gas \v{C}erenkov detector and pion rejector in the left spectrometer did not perform as well as the  gas \v{C}erenkov detector and preshower/shower detector in the right spectrometer in identifying pions. A tighter cut on momentum ($-4\% <\delta< 0\%$) was applied in the data analysis to avoid using the bad PMTs in the left gas \v{C}erenkov detector. The pion rejector was only used to estimate the corrections. The reconstructed photon energy spectrum from the data agreed with that from the simulation, as shown in Figure~\ref{fig:comp_rev}. For a consistency check, there were also some data recorded with both normal polarities and reversed polarities. The differences in yields were within 5\%, smaller than the systematic uncertainties (on the level of 10\%). 

\section{Results and Discussion} 
\label{results}
\begin{table*}[htbp] 
\caption{Differential cross sections for the $\gamma n \rightarrow \pi^- p$ process followed by the statistical and systematic uncertainties.} 
\begin{center} 
\begin{tabular}{|c|c|c|c|c|c|} 
\hline\hline 
$E_e$  & $E_\gamma$  & $\sqrt{s}$ & $\theta_{c.m.}$ & $(\frac{d\sigma}{d\Omega})_{c.m.}$  & $s^7 \frac{d\sigma}{dt}$  \\ 
(GeV) & (GeV) & (GeV) &($^\circ$) & ( $\mu$b/sr )& (10$^7$ nb $\cdot$ GeV$^{12}$) \\ 
\hline 
5.614 & 5.536  & 3.36  &  89.6 &   (4.22 $\pm$ 0.09 $\pm$ 0.42)$\times 10^{-4}$  
&  1.28 $\pm$ 0.03 $\pm$ 0.13 \\ 
      & 5.529  & 3.36  &  70.5 &   (1.05 $\pm$ 0.03 $\pm$ 0.12)$\times10^{-3}$   
&  3.17 $\pm$   0.08 $\pm$   0.35 \\ 
\hline 
4.236 & 4.158  & 2.95  &  89.8 &   (2.56 $\pm$   0.04 $\pm$ 0.26)$\times10^{-3}$   
&  1.71 $\pm$   0.03 $\pm$   0.17 \\ 
      & 4.157  & 2.95  &  69.8  &  (3.64 $\pm$   0.07 $\pm$ 0.40)$\times10^{-3}$   
&  2.43 $\pm$   0.05  $\pm$  0.27 \\ 
      & 4.141  & 2.94  &  50.1 &   (2.95 $\pm$   0.07 $\pm$ 0.32)$\times10^{-2}$   
& 19.3  $\pm$  0.45  $\pm$  2.13 \\ 
\hline 
3.400 & 3.321  & 2.67  &  90.1 &  (5.66 $\pm$   0.06  $\pm$  0.57)$\times10^{-3}$    
& 1.20  $\pm$  0.01  $\pm$  0.12 \\ 
      & 3.321  & 2.67  &  69.8 &  (1.50 $\pm$   0.02 $\pm$   0.16)$\times10^{-2}$    
& 3.19  $\pm$  0.03  $\pm$  0.35 \\ 
      & 3.322  & 2.67  &  49.8  & (6.63 $\pm$   0.09  $\pm$ 0.73)$\times10^{-2}$   
& 14.1  $\pm$  0.20  $\pm$   1.55 \\ 
      & 3.320  & 2.67  & 100.0  & (1.34 $\pm$  0.03 $\pm$  0.13)$\times10^{-2}$    
& 2.85   $\pm$  0.06  $\pm$   0.28 \\ 
      & 3.322  & 2.67  & 110.0  & (2.60 $\pm$ 0.04 $\pm$  0.26)$\times10^{-2}$    
& 5.53   $\pm$  0.09 $\pm$    0.55 \\ 
\hline 
2.561 & 2.481  & 2.36  &  89.9 &  (8.24  $\pm$ 0.10 $\pm$ 0.82)$\times10^{-2}$   
& 4.24 $\pm$   0.05 $\pm$   0.42 \\ 
      & 2.482  & 2.36  &  69.8 &   (6.18 $\pm$ 0.08 $\pm$  0.68)$\times10^{-2}$    
& 3.19 $\pm$   0.04 $\pm$    0.35 \\ 
      & 2.484  & 2.36  &  49.7 &  (9.06  $\pm$  0.02 $\pm$ 1.00)$\times10^{-2}$     
& 4.69 $\pm$   0.09 $\pm$   0.52 \\ 
\hline 
1.877 & 1.815  & 2.07  &  89.9 & (3.68 $\pm$  0.02 $\pm$  0.37)$\times10^{-1}$   
& 4.58 $\pm$   0.03 $\pm$    0.46 \\ 
      & 1.813  & 2.07  & 49.9 &(4.74  $\pm$ 0.07  $\pm$ 0.52)$\times10^{-1}$    
& 5.88 $\pm$   0.09 $\pm$    0.65 \\ 
\hline 
1.723 & 1.659  & 2.00  &  89.9 &(4.96  $\pm$ 0.02 $\pm$  0.50)$\times10^{-1}$   
&  4.20 $\pm$   0.02 $\pm$   0.42 \\ 
      & 1.660  & 2.00  &  69.9 &(6.35 $\pm$ 0.04 $\pm$  0.70)$\times10^{-1}$    
& 5.40 $\pm$   0.03 $\pm$    0.59 \\ 
      & 1.659  & 2.00  &  49.9 &(7.47  $\pm$ 0.06  $\pm$ 0.82)$\times10^{-1}$    
& 6.33 $\pm$   0.05 $\pm$    0.70 \\ 
\hline 
1.173 & 1.104  & 1.72  &  90.2 &(6.83 $\pm$  0.04 $\pm$  0.68)$\times10^{-1}$    
& 1.17 $\pm$   0.01 $\pm$   0.12 \\ 
      & 1.105  & 1.72  &  70.2 & 1.48 $\pm$  0.01  $\pm$ 0.16   
&  2.55 $\pm$   0.01 $\pm$   0.28 \\ 
      & 1.105  & 1.72  &  50.2 & 3.79 $\pm$  0.02 $\pm$  0.42 
& 6.53  $\pm$  0.04 $\pm$   0.72 \\ 
\hline \hline 
\end{tabular} 
\end{center} 
\label{cross_coin_tab} 
\end{table*} 
\begin{table*}[htbp] 
\caption{Differential cross sections for the $\gamma p \rightarrow \pi^+ n$ process followed by the statistical and systematic uncertainties.} 
\begin{center} 
\begin{tabular}{|c|c|c|c|c|c|} 
\hline\hline 
$E_e$  & $E_\gamma$  & $\sqrt{s}$ & $\theta_{c.m.}$ & $(\frac{d\sigma}{d\Omega})_{c.m.}$  & $s^7 \frac{d\sigma}{dt}$  \\ 
(GeV) & (GeV) & (GeV) &($^\circ$) & ( $\mu$b/sr )& (10$^7$ nb $\cdot$ GeV$^{12}$) \\ 
\hline 
5.614 & 5.535  & 3.36   & 89.8 &  (2.55 $\pm$  0.15 $\pm$  0.20)$\times 10^{-4}$  &  0.77  $\pm$  0.05 $\pm$   0.06 \\ 
      & 5.537  & 3.36  & 100.0 &  (2.44 $\pm$  0.11 $\pm$  0.19)$\times 10^{-4}$  &  0.74  $\pm$  0.03 $\pm$   0.06 \\ 
\hline 
4.236 & 4.156  & 2.95   & 89.9 &  (1.40  $\pm$ 0.03  $\pm$  0.11)$\times10^{-3}$  &  0.94 $\pm$   0.02 $\pm$   0.08 \\ 
      & 4.156  & 2.95   & 69.7 &  (1.79  $\pm$ 0.03 $\pm$  0.16)$\times10^{-3}$  &  1.19 $\pm$   0.02 $\pm$   0.11 \\ 
      & 4.156  & 2.95  & 100.0 &  (1.14 $\pm$   0.03  $\pm$   0.09)$\times10^{-3}$  &  0.76 $\pm$    0.02  $\pm$   0.06\\ 
\hline 
3.400 & 3.319  & 2.67   & 89.9 &  (3.67 $\pm$  0.05  $\pm$  0.29)$\times10^{-3}$  &  0.78 $\pm$   0.01 $\pm$   0.06 \\ 
      & 3.319  & 2.67   & 69.7 &  (1.78 $\pm$  0.01 $\pm$  0.16)$\times10^{-2}$  &  3.79 $\pm$  0.03 $\pm$   0.34 \\ 
      & 3.321  & 2.67   & 49.7 &  (1.58 $\pm$  0.01 $\pm$  0.14)$\times10^{-1}$  & 33.6 $\pm$   0.18 $\pm$   3.02 \\ 
      & 3.320  & 2.67  & 100.0 &  (8.02 $\pm$  0.11  $\pm$ 0.64)$\times10^{-3}$  &  1.71  $\pm$  0.02  $\pm$  0.14\\ 
      & 3.320  & 2.67  & 109.9 &  (9.51 $\pm$  0.23 $\pm$  0.76)$\times10^{-3}$  &  2.02 $\pm$   0.05 $\pm$   0.16 \\ 
\hline 
2.561 & 2.481  & 2.35  &  90.0 &  (5.88 $\pm$  0.05 $\pm$  0.47)$\times10^{-2}$  &  3.03 $\pm$   0.03  $\pm$   0.24 \\ 
      & 2.481  & 2.35   & 69.9 &  (1.01 $\pm$  0.01 $\pm$  0.09)$\times10^{-1}$  &  5.21  $\pm$  0.05 $\pm$   0.47 \\ 
      & 2.483  & 2.35   & 49.8 &  (3.29 $\pm$  0.03 $\pm$  0.30)$\times10^{-1}$  & 17.0  $\pm$  0.13 $\pm$    1.53 \\ 
\hline 
1.877 & 1.801  & 2.06  &  89.6 &  (2.42  $\pm$ 0.01  $\pm$  0.19)$\times10^{-1}$  &  2.91  $\pm$  0.01 $\pm$   0.23 \\ 
      & 1.805  & 2.07  &  49.6 &  (9.01 $\pm$  0.04 $\pm$  0.81)$\times10^{-1}$  & 11.0  $\pm$  0.05 $\pm$   0.99 \\ 
\hline 
1.723 & 1.647  & 1.99  &  89.6 &  (2.89 $\pm$  0.02 $\pm$  0.23)$\times10^{-1}$  &  2.38 $\pm$    0.01 $\pm$   0.19 \\ 
      & 1.648  & 1.99  &  69.5 &  (5.95 $\pm$  0.03 $\pm$  0.54)$\times10^{-1}$  &  4.91 $\pm$   0.03 $\pm$   0.44 \\ 
      & 1.650  & 1.99  &  49.6 &  1.15 $\pm$  0.01 $\pm$  0.10   & 9.49  $\pm$  0.05 $\pm$   0.85 \\ 
\hline 
1.173 & 1.097  & 1.71  &  90.0 &  1.35 $\pm$  0.01 $\pm$  0.11  &  2.26 $\pm$   0.01 $\pm$   0.18 \\ 
      & 1.098  & 1.72  &  70.0 &  3.13 $\pm$  0.01 $\pm$  0.28  &  5.27 $\pm$   0.02 $\pm$   0.47 \\ 
       
\hline \hline 
\end{tabular} 
\end{center} 
\label{cross_sing_tab} 
\end{table*} 
 
\begin{figure}[htbp] 
\centerline{\includegraphics[bb=15 150 530 650,scale=0.45]{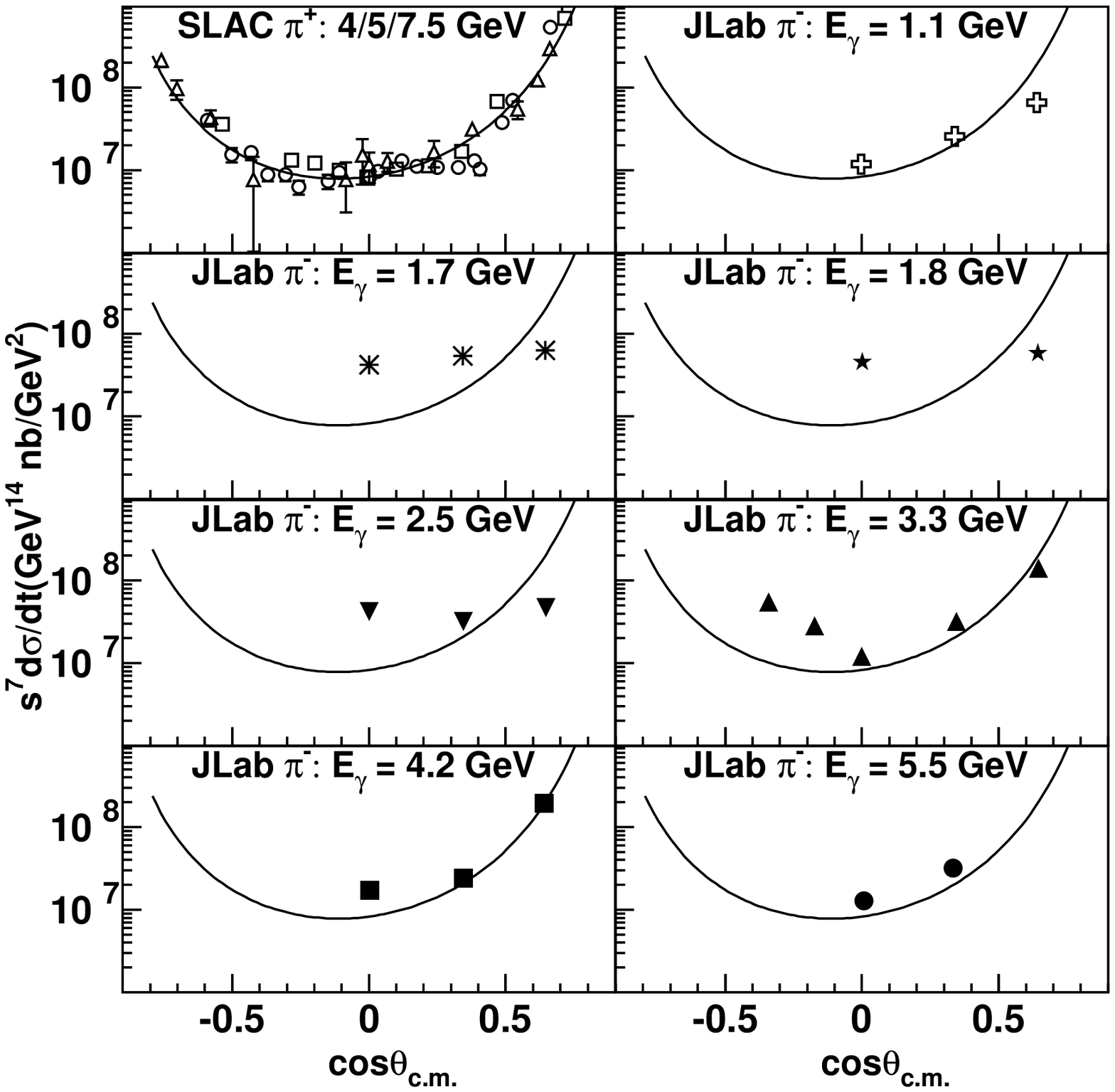}} 
\caption{Angular distributions from JLab E94-104 for the $\gamma n \rightarrow \pi^- p$ process, as well as those from the SLAC data~\cite{anderson76} for the $\gamma p \rightarrow \pi^+ n$ process at photon energy of 4 GeV (open squares), 5 GeV (open circles), and 7.5 GeV (open triangles). The curve in each panel is the empirical fit of SLAC data: $0.828e7 \cdot (1-{\rm cos}\theta_{\rm c.m.})^{-5}\cdot(1+{\rm cos}\theta_{\rm c.m.})^{-4}$.  } +
\label{angular_1} 
\end{figure}

\begin{figure}[htbp] 
\centerline{\includegraphics[bb=15 150 530 650,scale=0.45]{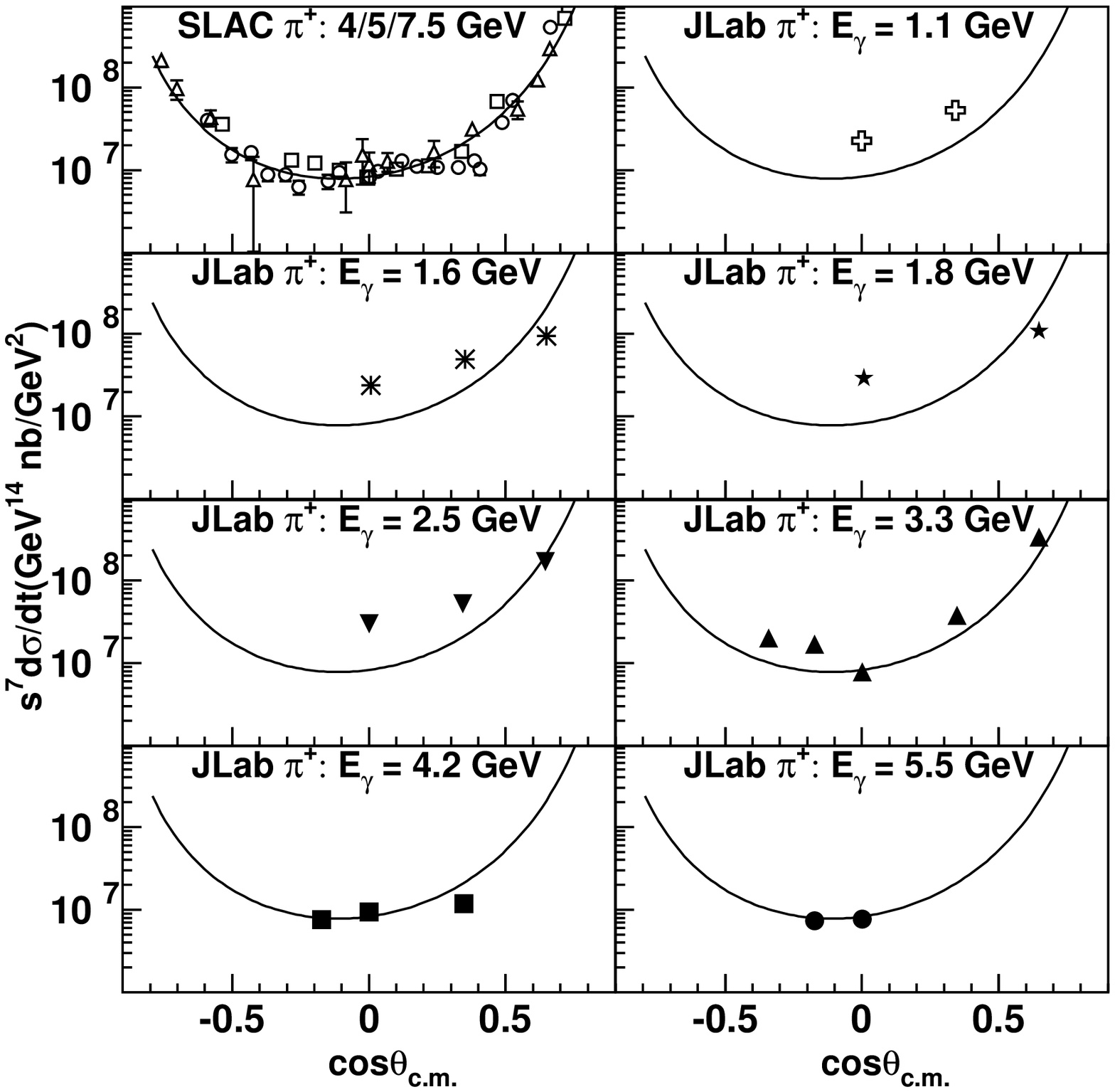}} 
\caption{Angular distributions from JLab E94-104 for the $\gamma p \rightarrow \pi^+ n$ process, as well as those from the SLAC data~\cite{anderson76} for the $\gamma p \rightarrow \pi^+ n$ process at photon energy of 4 GeV (open squares), 5 GeV (open circles), and 7.5 GeV (open triangles). The curve in each panel is the empirical fit of SLAC data: $0.828e7 \cdot (1-{\rm cos}\theta_{\rm c.m.})^{-5}\cdot(1+{\rm cos}\theta_{\rm c.m.})^{-4}$.  } 
\label{angular_2} 
\end{figure}

The differential cross sections $d\sigma/d\Omega$ and $s^7d\sigma/dt$ extracted from JLab experiment E94-104 are shown in Table~\ref{cross_coin_tab} and Table~\ref{cross_sing_tab} for different beam energies and pion center-of-mass angles. The published results at 90$^\circ$~\cite{prl_90} are also updated here. The angular distributions for all the energies are plotted in Figure~\ref{angular_1} and Figure~\ref{angular_2}. Also plotted are the SLAC $\pi^+$ data at $E_\gamma=4,5,7.5$ GeV~\cite{anderson76}. Both the $\pi^+$ and $\pi^-$ data at comparable energies, i.e. $E_\gamma=4.2,5.5$ GeV, are consistent with the fit of the high energy SLAC data.  
 
 
It is worth mentioning that the pQCD calculations involving gluon self-coupling~\cite{farrar91} cannot reproduce the angular distribution of the SLAC data, especially at the backward angles. This discrepancy may be due to the relatively low values of $s$, $|t|$ and $u$. The main contamination of their leading-twist predictions came from the $t$-channel meson resonances at forward angles, and from the $u$-channel baryon resonances at backward angles. 
 
\subsection{Comparison with the World Data}
\begin{figure}[htbp] 
\centerline{\includegraphics*[bb=20 150 550 670, scale=0.45]  {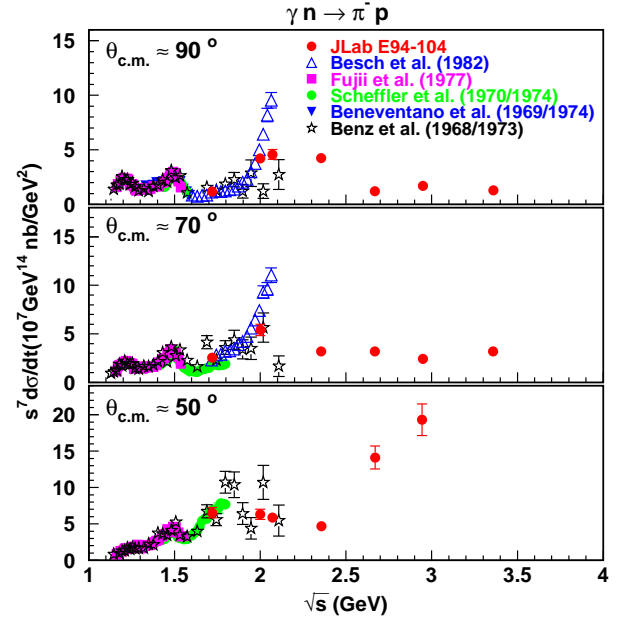}} 
\caption{(Color online). Scaled differential cross section $s^7\frac{d\sigma}{dt}$ versus center-of-mass energy $\sqrt{s}$ for the $\gamma n \rightarrow \pi^- p$ process from JLab E94-104 and previous world data~\cite{world_data,besch,world_ref}. 
The open triangles in the upper panel are averaged from the Besch {\it et al.} data~\cite{besch} at $\theta_{c.m.}=85^\circ$ and 95$^\circ$, while open triangles in the middle panel are averaged from those at $\theta_{c.m.}=65^\circ$ and 75$^\circ$. The  Fujii {\it et al.} and  Scheffler {\it et al.} data~\cite{world_ref} in the middle panel were taken at $\theta_{c.m.}=75^\circ$. } 
\label{cross_coin} 
\end{figure}

\begin{figure}[htbp] 
\centerline{\includegraphics*[bb=20 150 550 670,scale=0.45]{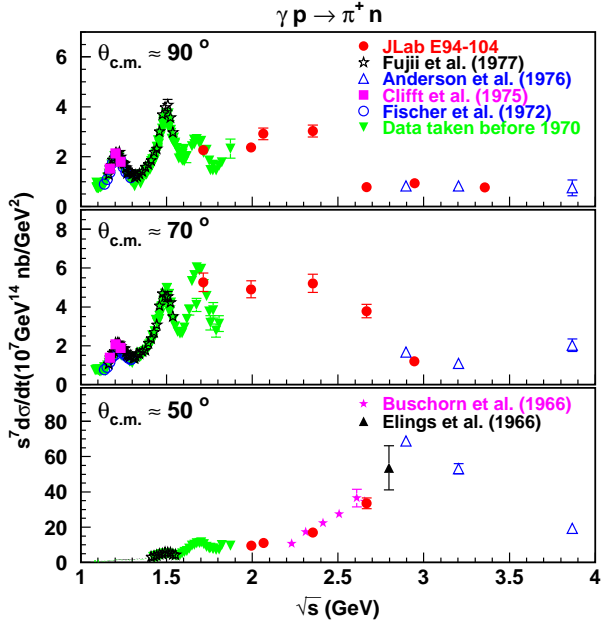}} 
\caption{(Color online). Scaled differential cross section $s^7\frac{d\sigma}{dt}$ versus center-of-mass energy $\sqrt{s}$ for the $\gamma p \rightarrow \pi^+ n$ process from JLab E94-104 and previous world data~\cite{world_data,anderson76,world_ref}.} 
\label{cross_sing} 
\end{figure}

As shown in Figure~\ref{cross_coin} and Figure~\ref{cross_sing}, the data from JLab experiment E94-104 (in solid circles) extended the single pion photoproduction measurements at several GeV~\cite{world_data,anderson76,besch,world_ref}, by spanning the resonance region and the scaling region. The differential cross sections of the $\gamma n \rightarrow \pi^- p$ process with $\sqrt{s}$ greater than 2.2 GeV were measured for the first time. The uncertainty in $\sqrt{s}$ due to the 100 MeV photon energy window ranges from 0.05 to 0.03 GeV as the beam energy increases  
 
The data agree within uncertainties with the world data in the overlapping energy region, except with the Besch {\it et al.} data~\cite{besch} (open triangles). The Besch {\it et al.} data from Bonn suggest a very sharp peak in the scaled cross section for the $\gamma n \rightarrow \pi^- p$ process with $\sqrt{s}$ around 2.0 GeV. Our data confirm the scaled cross section enhancement around that region, but the peak is much less pronounced.  
We do not know the origin of this discrepancy exactly, though our momentum resolution (0.02\%) is much better than that of Besch {\it et al.} (4\%). The broad structure suggested by our data is seen in $\pi^0$ photoproduction channel as well~\cite{world_data}. Similar broad structure was also seen in the $\pi^{+}$ and $\pi^0$ channels from the preliminary JLab CLAS (CEBAF Large Acceptance Spectrometer) results~\cite{CLAS}. 
 
\subsection{Scaling at High Energy}

Based on the constituent counting rule, the differential cross section $d\sigma/dt$ at a fixed center-of-mass angle for the $\gamma n \rightarrow \pi^{-} p$ and $\gamma p \rightarrow \pi^{+} n$ processes is predicted to scale as $s^{-7}$.
For both $\pi^-$ and $\pi^+$ photoproduction processes, the data with $\sqrt{s}> 2.7$ GeV at $\theta_{c.m.}=90^\circ$ and $\sqrt{s}> 3.0$ GeV at $\theta_{c.m.}=70^\circ$ indicate the scaling behavior of the differential cross section predicted by the constituent counting rule. The fitted power of $\frac{1}{s}$ from the three data points in this region was 6.9$\pm$0.2 for the   
$\gamma n \rightarrow \pi^{-} p$ process and 7.1$\pm$0.2 for the   
$\gamma p \rightarrow \pi^{+} p$ process, consistent with the prediction of 7. 
This may have some theoretical implications, for example the validity of quark-gluon degrees of freedom and the freezing of the running strong coupling constant at several GeV. 



There is no sign of $s^{-7}$ scaling for the data at $\theta_{c.m.}=50^\circ$ up to center-of-mass energy of 3.0 GeV for the $\pi^-$ case and 3.9 GeV for the $\pi^+$ case. This is not surprising since the deuteron photodisintegration data~\cite{schulte01,hallb04,schulte_th} at forward angles do not scale at as low energies as those at $90^\circ$. The deuteron photodisintegration data at $\theta_{c.m.}=53^\circ$ seem to scale when the photon energy is greater than 3 GeV while the data at $\theta_{c.m.}=90^\circ$ scale when the photon energy is greater than 1 GeV. The corresponding center-of-mass energies are 3.8 GeV and 2.7 GeV respectively. 
 
\begin{figure}[htbp] 
\centerline{\includegraphics*[bb=40 200 510 670, scale=0.45]  {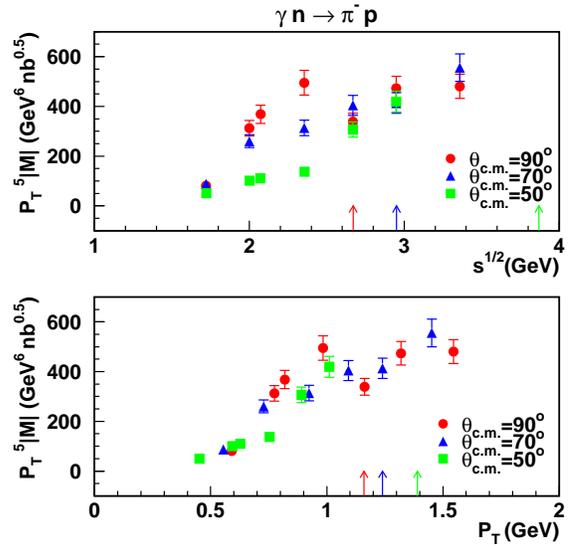}} 
\caption{(Color online). Scaled amplitude $P_T^5|M|$ versus center-of-mass energy $\sqrt{s}$ and transverse momentum $P_T$ for the $\gamma n \rightarrow \pi^- p$ process. All the data points came from JLab E94-104.  The arrows indicate the position below which the data do not scale for either $\pi^+$ or $\pi^-$ photoproduction.} 
\label{amp_pt_coin} 
\end{figure} 
\begin{figure}[htbp] 
\centerline{\includegraphics*[bb=40 200 510 670, scale=0.45]  {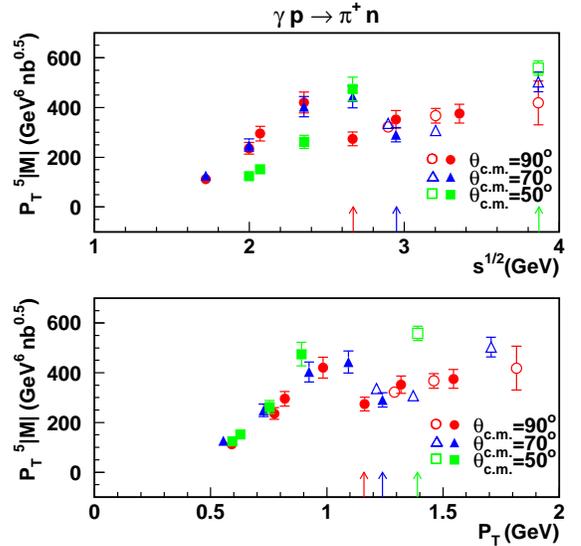}} 
\caption{(Color online). Scaled amplitude $P_T^5|M|$ versus center-of-mass energy $\sqrt{s}$ and transverse momentum $P_T$ for the $\gamma p \rightarrow \pi^+ n$ process. The data points in solid circles came from JLab E94-104, while those in open triangles are SLAC data~\cite{anderson76}. The arrows indicate the position below which the data do not scale for either $\pi^+$ or $\pi^-$ photoproduction.} 
\label{amp_pt_sing} 
\end{figure}

 The scaled invariant amplitude $P_T^5 |M|$ is plotted in Figure~\ref{amp_pt_coin} and Figure~\ref{amp_pt_sing} against center-of-mass energy $\sqrt{s}$ and transverse momentum $P_T$, similar to what was performed for neutral pion photoproduction on the deuteron~\cite{brodsky01}. The invariant amplitude $M$ was calculated from the differential cross section by using 
\begin{equation} 
|M|=4(s-m_N^2)\sqrt{\pi \frac{d\sigma}{dt}(\gamma N \rightarrow \pi N)}, 
\end{equation} 
and the transverse momentum was calculated by using $P_T=|\vec{p}_\pi|\sin \theta_\pi$ from the pion momentum $\vec{p}_\pi$ and scattering angle $\theta_\pi$. 
 The scaled amplitude $P_T^5 |M|$ is plotted here for the $\gamma N
\rightarrow \pi^{\pm} N$ process, while $P_T^{11} |M|$ was plotted in reference~\cite{brodsky01} for the $\gamma d \rightarrow \pi^0 d$ process. The scaling power of $P_T$ can be estimated by dimensional counting. Since the arrows in different colors (indicating the possible onset of scaling for different angles) are closer in terms of transverse momentum than in terms of the center-of-mass energy, the transverse momentum may be a better choice to describe the scaling onset than the center-of-mass energy, which was also stated previously~\cite{hallb04b,RNA,calson97}. The photoproduction data seem to reach the scaling region when the transverse momentum is around 1.2 GeV/c. As a comparison, 
the deuteron photodisintegration data start to exhibit scaling when proton transverse momentum ranges from 1.0 to 1.3 GeV/c at proton center-of-mass angle between  $30^\circ$ to $150^\circ$, except that the proton transverse momentum threshold ranges from 0.6 to 0.7 GeV/c at center-of-mass angle of $45^\circ$, $135^\circ$ and $145^\circ$~\cite{hallb04b}.
Another interesting observation is that the scaled amplitude has much less dependence on center-of-mass angle than do the scaled cross sections.
 
\begin{figure}[htbp] 
\centerline{\includegraphics*[bb=14 200 535 670,width=8cm,height=5cm]{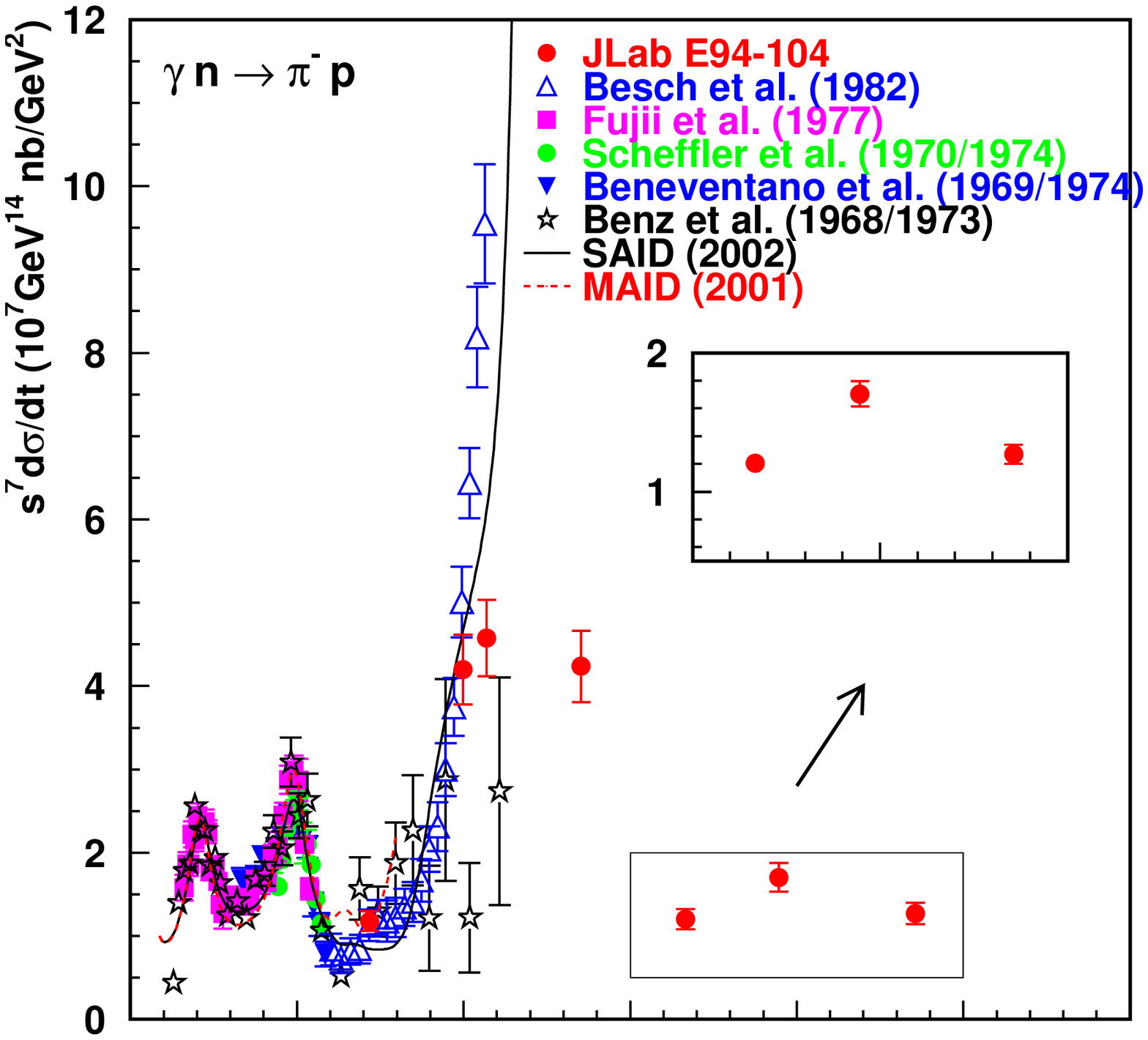}} 
\centerline{\includegraphics*[bb=14 157 535 643,width=8cm,height=5cm]{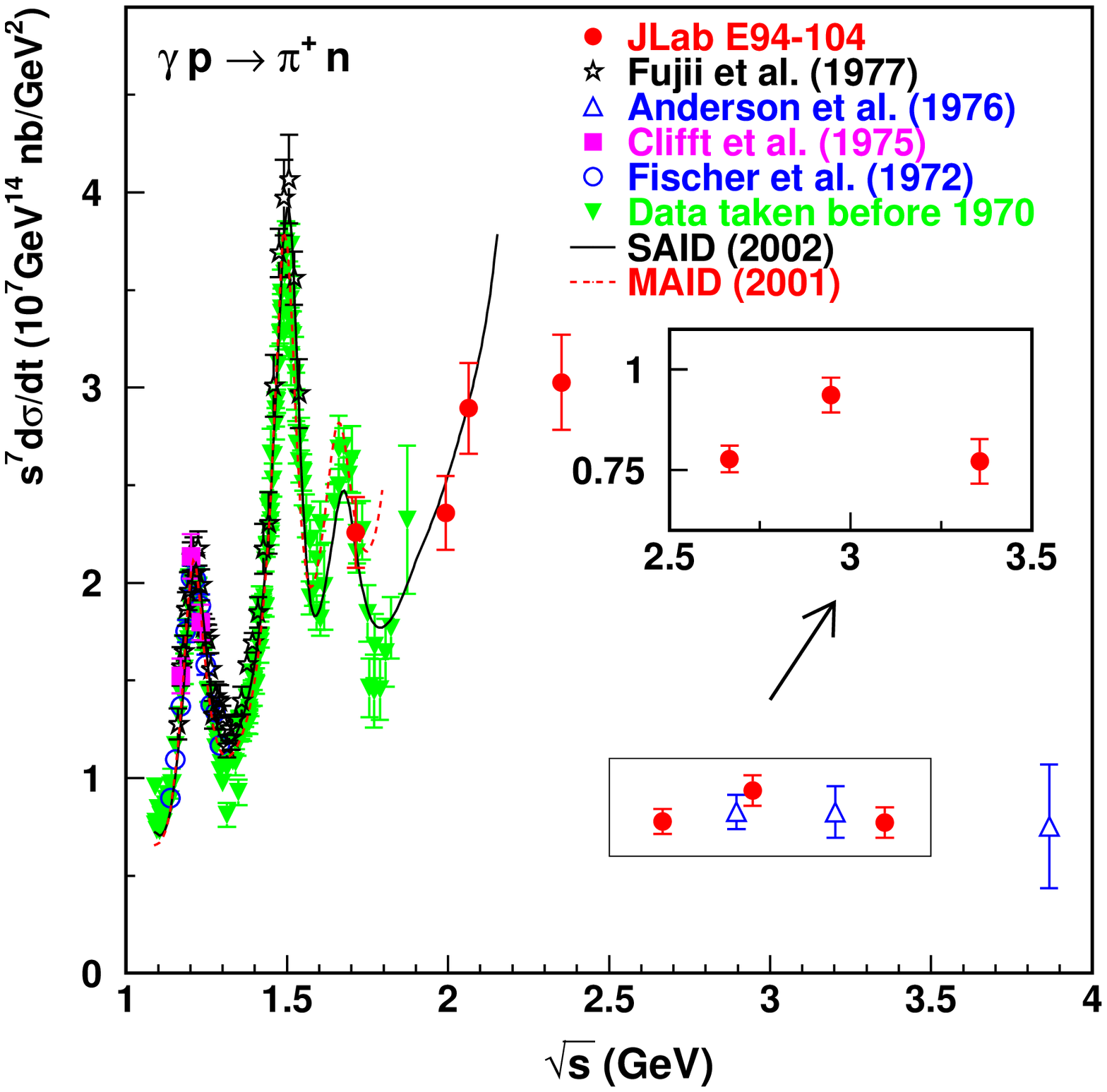}} 
\caption{(Color online). Scaled differential cross section $s^7 \frac{d\sigma}{dt}$ versus center-of-mass energy $\sqrt{s}$ for the $\gamma n \rightarrow \pi^- p$ process (upper panel) and $\gamma p \rightarrow \pi^+ n$ process (lower panel) at $\theta_{c.m.}=90^\circ$ from JLab E94-104 and previous world data~\cite{world_data,anderson76,besch,world_ref}. The error bars in the insets include only point-to-point uncertainties to highlight the possible substructure of scaling.  The solid line was obtained from the recent partial-wave analysis of the single pion photoproduction data up to $E_\gamma=2$ GeV~\cite{said}, while the dashed line was obtained from the MAID analysis up to $E_\gamma=1.25$ GeV~\cite{maid}} 
\label{pi_oscillation} 
\end{figure} 
 
\subsection{Possible Substructure of Scaling}
As shown in Figure~\ref{pi_oscillation}~\cite{prl_90}, though the data with current total uncertainties are consistent with the global scaling behavior, the data suggest some substructure for both processes: the point around 3.0 GeV is higher than those at 2.7 GeV and 3.4 GeV. This might be a hint of possible oscillations around the scaling value, similar to what was observed in the $pp$ elastic scattering data~\cite{hendry}.  Future measurements planned at JLab~\cite{E02-010}, with finer binning in beam energy, are essential for the confirmation of such oscillatory scaling behavior. 
 
The substructure or the possible oscillatory scaling behavior in pion photoproduction may arise from the same mechanism as in the case of $pp$ scattering, i.e. the interference between the long-distance (due to Landshoff diagrams) amplitude and short-distance amplitude~\cite{ralston}, or the interference between resonances with a pQCD background~\cite{brodsky_de}. But it may also be due to other mechanisms, such as high energy resonances around 3 GeV~\cite{pdg}, the interference between the amplitudes associated with different helicity changes based on the generalized counting rule~\cite{ji03,brodsky03,dutta04}, or the breaking of the local quark-hadron duality above the resonance region~\cite{zhao03}.

The generalized constituent counting rule~\cite{ji03} was derived for 
hard exclusive processes involving parton orbital momentum and hadron 
helicity flip. It predicts that 
$M(\gamma N_\uparrow \rightarrow \pi N_\downarrow) \sim s^{-3}$ 
and $d\sigma/dt(\gamma N_\uparrow \rightarrow \pi N_\downarrow) \sim s^{-8}$ 
for the helicity-flip case, while 
$M(\gamma N_\uparrow \rightarrow \pi N_\uparrow)\sim s^{-5/2}$ 
and $d\sigma/dt(\gamma N_\uparrow \rightarrow \pi N_\uparrow) \sim s^{-7}$ 
for the helicity conserving case. By including the amplitudes with helicity changes and their interference with the long-distance amplitudes, one is able to reproduce~\cite{dutta04} the anomalous oscillations of the differential cross section $d\sigma /dt$ around the scaling value in $pp$ elastic scattering~\cite{hendry} especially in the low energy region ($s<10$ GeV$^2$), as well as the very large spin-spin correlation~\cite{crabb}. This may be also true for pion photoproduction processes.
 
The locality of quark-hadron duality means that the local averages of physical variables measured in the resonance region are equal to those measured in the deep-inelastic or scaling region. The locality of quark-hadron duality can be realized in a simple model of a composite system with two spinless charged constituents described by harmonic oscillator wave functions with principal quantum number $N$ and orbital angular momentum $L$ ($\leq N$). The destructive interference between the high density of overlapping resonances leads to the smooth scaling behavior at high energies. But for medium energies, the locality of quark-hadron duality may break down and a sizable oscillation around the scaling value may appear above the resonance region due to the orbital angular momentum dependence of the resonances~\cite{zhao03}. In this mechanism, the energy increase reduces the oscillation amplitude, and the $Q^2$ dependence may be nontrivial. If a subset of resonances are relatively suppressed at large $Q^2$, there will be significant shifts in the position and magnitude of oscillations. The deviation pattern produced by the resonance degeneracy breaking requires no simple periodicity. The experimental data can thus distinguish this mechanism from others. 

\subsection{Cross Section Enhancement around 2.2 GeV}
Another interesting feature of the differential cross section is an apparent enhancement with a sharp drop for both channels of the charged pion photoproduction at $\theta_{c.m.}=90^\circ$, at a center-of-mass energy ranging approximately from 1.8 to 2.5 GeV, as shown in Figure~\ref{pi_oscillation}. Note that it is the scaled cross sections $s^7d\sigma/dt$ that are plotted. The drop is even faster in terms of non-scaled cross section. A similar cross section enhancement was also observed in neutral pion photoproduction~\cite{world_data,CLAS}. But the enhancement patterns are different for the kinematics at $\theta_{c.m.}=70^\circ$ and $\theta_{c.m.}=50^\circ$.  

Some speculation can be made about the enhancement. 
It might be due to the known baryon resonances around this energy, for example $G_{17}$(2190), $H_{17}$(2220) and $G_{19}$(2250)~\cite{pdg},  just as is the case at lower energies.  
It might relate to some missing resonances~\cite{capstick}, which were predicted by the constituent quark model but have not been seen experimentally. The value of center-of-mass energy hints that the enhancement might be associated with the strangeness production threshold, which is around 2 GeV to produce a $\phi$ meson of mass 1 GeV. 
The resonances at strangeness and charm production thresholds were assumed in an approach to explain the strong spin-spin correlation and oscillatory scaling in elastic $pp$ scattering~\cite{brodsky_de}.  It is worthwhile to mention that a broad bump near 2.2 GeV appears in the $\pi^- p$ total cross section, while it is not clear in $\pi^- p$ elastic cross section~\cite{pdg}. 
 
 \subsection{Exclusive Charged Pion Ratio}


\begin{figure}[htbp] 
\centerline{\includegraphics*[bb=15 162 570 670,scale=0.4]{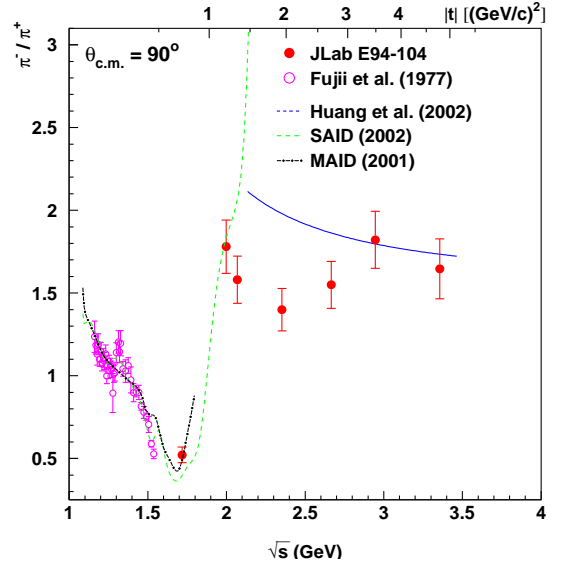}} 
\caption{(Color online). Exclusive charged pion ratio $\frac{d\sigma/dt(\gamma n \rightarrow \pi^{-} p)}{d\sigma/dt(\gamma p \rightarrow \pi^{+} n)}$ versus center-of-mass energy $\sqrt{s}$ and momentum transfer squre $|t|$ at pion center-of-mass angle $\theta_{c.m.}=90^\circ$ from JLab E94-104 and previous world data~\cite{world_ref}, together with the SAID~\cite{said}, MAID~\cite{maid} and one-hard-gluon-exchange calculation~\cite{huang1,huang2}.   } 
\label{fig:ratio} 
\end{figure} 

 One can form the exclusive charged pion ratio $\frac{d\sigma/dt(\gamma n \rightarrow \pi^{-} p)}{d\sigma/dt(\gamma p \rightarrow \pi^{+} n)}$ based on the E94-104 data. As shown in Figure~\ref{fig:ratio}, the exclusive charged pion ratio has some oscillations at low energies due to the isospin dependence of the resonances, which can be described by the SAID~\cite{said} and MAID~\cite{maid} calculations available at low energies. The big jump around 2 GeV might be associated with the isospin-dependent resonances nearby, or with the strangeness production threshold (around 2 GeV for $\phi$ production). 
The lowest order (leading-twist) calculation based on one-hard-gluon-exchange diagrams~\cite{huang1,huang2}, which is only valid at high energies, predicts a smooth and simple behavior of    
$
\frac{ d\sigma/dt(\gamma n \rightarrow \pi^-p) }{d\sigma/dt(\gamma p \rightarrow \pi^+n) } \simeq  ( \frac{ue_d+se_u}{ue_u+se_d} )^2 \, 
$
after the nonperturbative components represented by the form factors are divided out in the ratio. 
The theoretical prediction seems to agree with the $\theta_{c.m.}=90^\circ$ data at the two highest energies.  
 
\begin{figure}[htbp] 
\centerline{\includegraphics*[bb=15 133 540 670,scale=0.45]{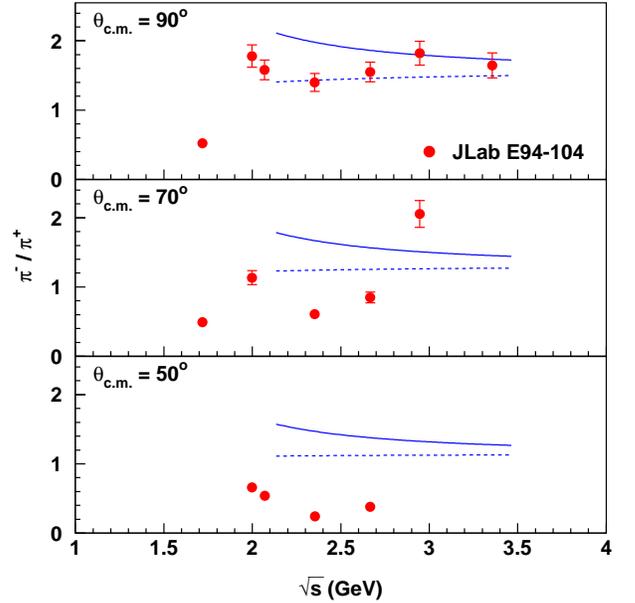}} 
\caption{(Color online). Exclusive charged pion ratio $\frac{d\sigma/dt(\gamma n \rightarrow \pi^{-} p)}{d\sigma/dt(\gamma p \rightarrow \pi^{+} p)}$ versus center-of-mass energy $\sqrt{s}$ at different pion center-of-mass angles from JLab E94-104. The solid curve is calculated by using Equation~\ref{eq:ratio}, while the dashed one considers the nucleon mass by using Equation~\ref{eq:ident}. } 
\label{ratio_all} 
\end{figure} 
 
The exclusive charged pion ratio was also calculated to higher orders (twist-2 and twist-3) within the handbag mechanism considering both quark helicity flip and non-flip~\cite{huang3}. The more precise approach led to the same result as the leading-twist prediction when $|C_2^P| >> |C_3^P|$. The invariant functions $|C_i^P| \ (i=1,4)$ are the coefficients for the four gauge invariant covariants into which the meson photoproduction amplitudes can be decomposed. Both $|C_2^P|$ and $|C_3^P|$ contribute only to the quark helicity conserving amplitudes while $|C_1^P|$ and $|C_4^P|$ generate quark helicity flips. The charged pion ratio will become infinity at large $s$ if $|C_3^P|$ is dominant, which is clearly not supported by our data. 

The calculation was for massless particle. The effects due to nucleon mass may be important at an energy scale of few GeV, as estimated by the difference between the solid and dashed curves in Figure~\ref{ratio_all}. 
The calculation shown by the dashed curve considers the nucleon mass by using the identification~\cite{diehl} of the Mandelstam variables with the experimental ones ($s_{\rm exp}$, $t_{\rm exp}$, $u_{\rm exp}$)  
\begin{equation} 
s=s_{\rm exp}-m_p^2 ;  t=t_{\rm exp} ;  u=u_{\rm exp}-m_p^2,  
\label{eq:ident} 
\end{equation} 
so that $s+t+u \sim 0$. Figure~\ref{ratio_all} also shows that the agreement at forward angles is not as good as the case at $90^\circ$, which might be related to the relatively lower momentum transfer. Measurements at higher energies are necessary to fully check the theoretical approach, though the current comparison between prediction and data seems to suggest that the handbag mechanism is at work for the pion photoproduction processes with dominant quark helicity non-flip amplitudes and $|C_2^p| >> |C_3^p|$.  

\ \\ 

\section{Outlook} 
As mentioned above, a new experiment~\cite{E02-010} was proposed in JLab Hall A to measure the $\gamma n \rightarrow \pi^- p $ process with deuterium and carbon targets, as well as the $\gamma p \rightarrow \pi^+ n $ process with a hydrogen target. With very fine steps in center-of-mass energy, approximately 0.07 GeV, the new experiment will be able to elucidate more details about the possible substructure of the scaling behavior. In addition, the nuclear transparency of carbon in the pion photoproduction process will be measured for the first time. This should enable us to test some theoretical predictions such as the nuclear filtering effect and color transparency. The latter was suggested by the helium transparency measurement in E94-104~\cite{prc_he}. 
 
Since the cross section decreases relatively slowly as energy increases, the measurements for single pion photoproduction processes can be greatly extended with the JLab 12 GeV upgrade. Since the charm production threshold will be crossed, one would be able to investigate the resonance at this threshold that was assumed in an approach to explain the anomalies in $pp$ scattering~\cite{brodsky_de}. In addition, one would be able to further test the scaling and charged pion ratio predictions, especially at forward angles. 

\section{Acknowledgments} 
We acknowledge the outstanding support of JLab Hall A technical staff and 
Accelerator Division in accomplishing this experiment. 
We thank R.~B.~Wiringa and H.~Arenh$\rm \ddot{o}$vel for calculating the  
momentum distribution of the neutron in the deuterium target.  
We thank Z.~Chai for providing the code to apply R-function cut on acceptance. 
We also thank S. Brodsky, T.~W.~Donnelly, H.~W.~Huang, P.~Jain, P. Kroll, J.~Ralston, C.~D.~Roberts, P.~Rossi and G.~de~Teramond for helpful discussions. 
This work was supported in part by the U.~S.~Department of Energy, DOE/EPSCoR, 
the U.~S.~National Science Foundation,  
the Ministero dell'Universit\`{a} e della Ricerca 
Scientifica e Tecnlogica (Murst), 
the French Commissariat \`{a} l'\'{E}nergie Atomique, 
Centre National de la 
Recherche Scientifique (CNRS) and the Italian Istituto Nazionale di Fisica 
Nucleare (INFN). 
This work was supported by DOE contract DE-AC05-84ER40150 
under which the Southeastern Universities Research Association 
(SURA) operates the Thomas Jefferson National Accelerator Facility. 
 
 
\end{document}